\documentclass[preprint,12pt]{elsarticle}

\usepackage{amssymb}

\usepackage{lipsum}
\usepackage{multicol}

\usepackage{amsmath,amsfonts,amssymb,graphicx,color,titlesec}
\usepackage[font=footnotesize]{caption}

\usepackage{caption}
\usepackage{subcaption} 

\usepackage{comment}

\usepackage{tikz}
\usetikzlibrary{arrows}
\usetikzlibrary{fit,positioning,hobby}
\usetikzlibrary{decorations.text}
\usetikzlibrary{decorations.pathreplacing}

\usepackage{centernot}
\usepackage{stmaryrd}
\usepackage{xfrac}

\usepackage{booktabs}

\usepackage{hyperref}

\definecolor{light-purple}{HTML}{AB47BC}
\definecolor{medium-purple}{HTML}{9c27b0}
\definecolor{dark-purple}{HTML}{7b1fa2}
\definecolor{grey}{HTML}{b8b8b8}

\newcommand{\dd}{{\rm d}}
\newcommand{\dt}{{\rm d}t}

\journal{
\empty}

\begin{document}

\begin{frontmatter}



\title{A story of viral co-infection, co-transmission and co-feeding in ticks: how to compute an invasion reproduction number}


\author[inst1,inst2]{Giulia Belluccini}

\affiliation[inst1]{organization={T-6, Theoretical Biology
and Biophysics},
            addressline={Los Alamos National Laboratory}, 
            city={Los Alamos},
            postcode={87545}, 
            state={NM},
            country={USA}}

\affiliation[inst2]{organization={School of Mathematics},
            addressline={University of Leeds}, 
            city={Leeds},
            postcode={LS2 9JT}, 
            country={UK}}

\author[inst1]{Qianying Lin}

\author[inst2]{Bevelynn Williams}

\author[inst3]{Yijun Lou}

\author[inst4]{Zati Vatansever}

\author[inst2]{Mart\'in L\'opez-Garc\'ia}

\author[inst2]{Grant Lythe}

\author[inst1]{Thomas Leitner}

\author[inst1]{Ethan Romero-Severson}

\author[inst1]{Carmen Molina-Par\'is}

\affiliation[inst3]{organization={Department of Applied Mathematics},
            addressline={Hong Kong Polytechnic University}, 
            city={Hong Kong SAR},
            country={ China }}

\affiliation[inst4]{organization={Department of Parasitology},
            addressline={Faculty of Veterinary Medicine, Kafkas University}, 
            city={Kars},
            country={Turkey}}

\begin{abstract}

With a single circulating vector-borne virus, the basic reproduction number incorporates contributions from tick-to-tick (co-feeding), tick-to-host and host-to-tick transmission routes. With two different circulating  vector-borne viral strains, resident and invasive, and under the assumption that co-feeding is the only transmission route in a tick population, the invasion reproduction number depends on whether the model system of ordinary differential equations possesses the property of neutrality. We show that a simple model, with two populations of ticks infected with one strain, resident or invasive, and one population of co-infected ticks, does not have Alizon's neutrality property. We present model alternatives that are capable of representing the invasion potential of a novel strain by including populations of ticks dually infected with the same strain. The invasion reproduction number is analysed with the next-generation method and via numerical simulations.

\end{abstract}



\begin{highlights}

\item We introduce a mathematical
model of a single vector-borne virus
in a population of ticks and hosts,
with three different transmission
routes, and derive its basic reproduction number.

\item 
We study the dynamics of two different co-circulating viruses, or viral strains, in a tick population making use of a classic co-infection model. 

\item 
After performing an invasion analysis, we
compute the invasion reproduction number, 
explain the issue of its
non-neutrality, and propose five  neutral alternatives.

\item
We conclude the paper with a summary of our proposals, their applicability and limitations. 

\end{highlights}


\end{frontmatter}



\section{Introduction}
\label{sec:introduction}

Co-infection of a single host by at least two distinct viruses provides an opportunity for viruses to exchange genetic information through genomic reassortment or 
recombination~\cite{mcdonald2016reassortment,perez2015recombination}.
In fact, entirely novel pathogenic viruses have emerged from reassortment events of less pathogenic parents in 
nature~\cite{negredo2021fatal,gerrard2004ngari,cline2011increased}.
Co-infection can be thought of as the rate-limiting step in the sudden emergence of genetically distant variants of existing human pathogens such as influenza, SARS-CoV-2, and Crimean Congo Hemorrhagic Fever Virus (CCHFV).  
Therefore, understanding the dynamics of co-infection in common host species, {\em e.g.,} arthropods (ticks or mosquitoes), is essential to study the emergence and re-emergence of both new and old human pathogens. 

\begin{figure}[htp!]
\centering
 \begin{tabular}{cc}
 {\bf A)} & {\bf B)}
 \\
 \includegraphics[scale=0.35]{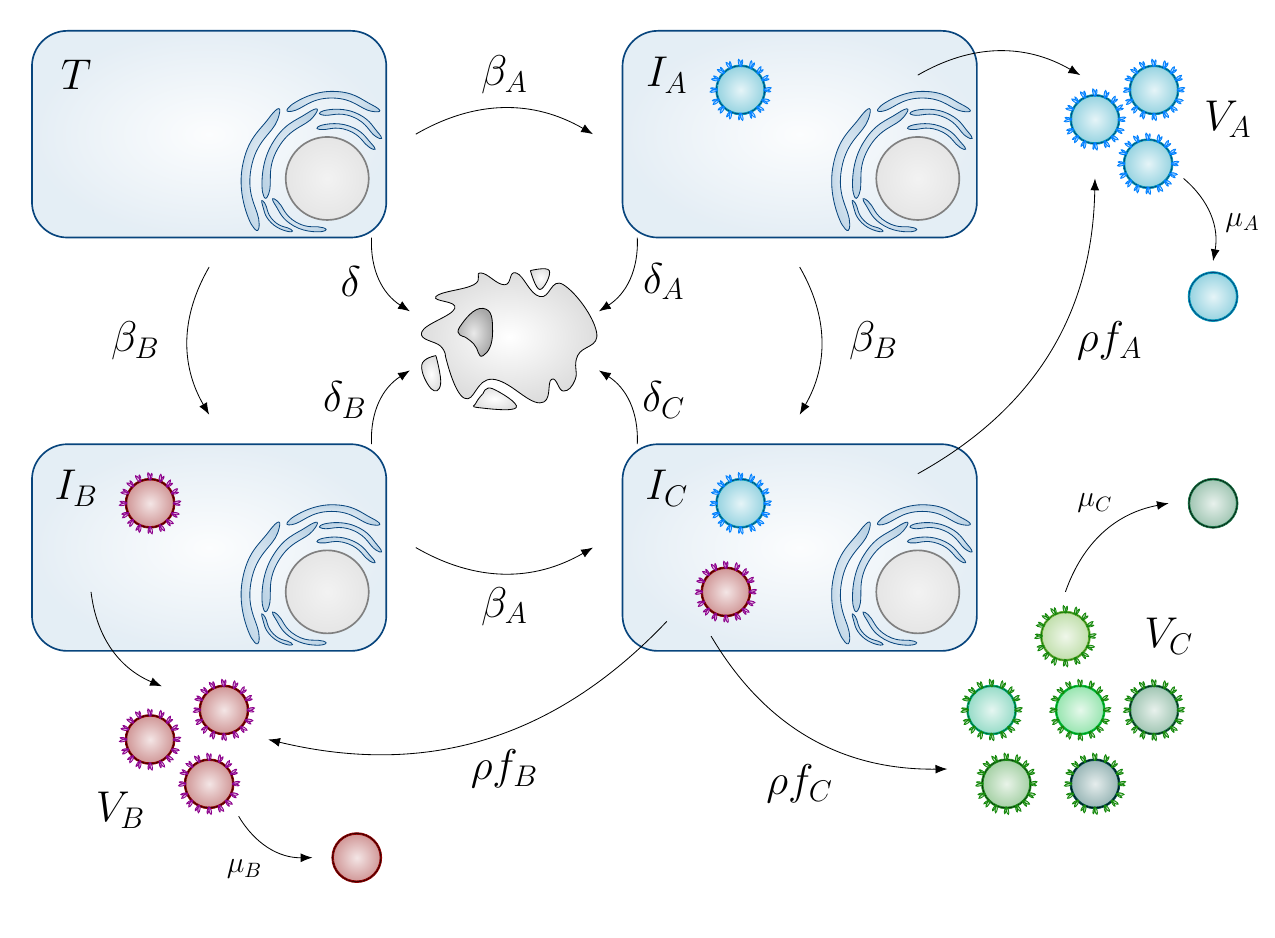} &
 \hspace{1cm}
 \includegraphics[scale=0.35]{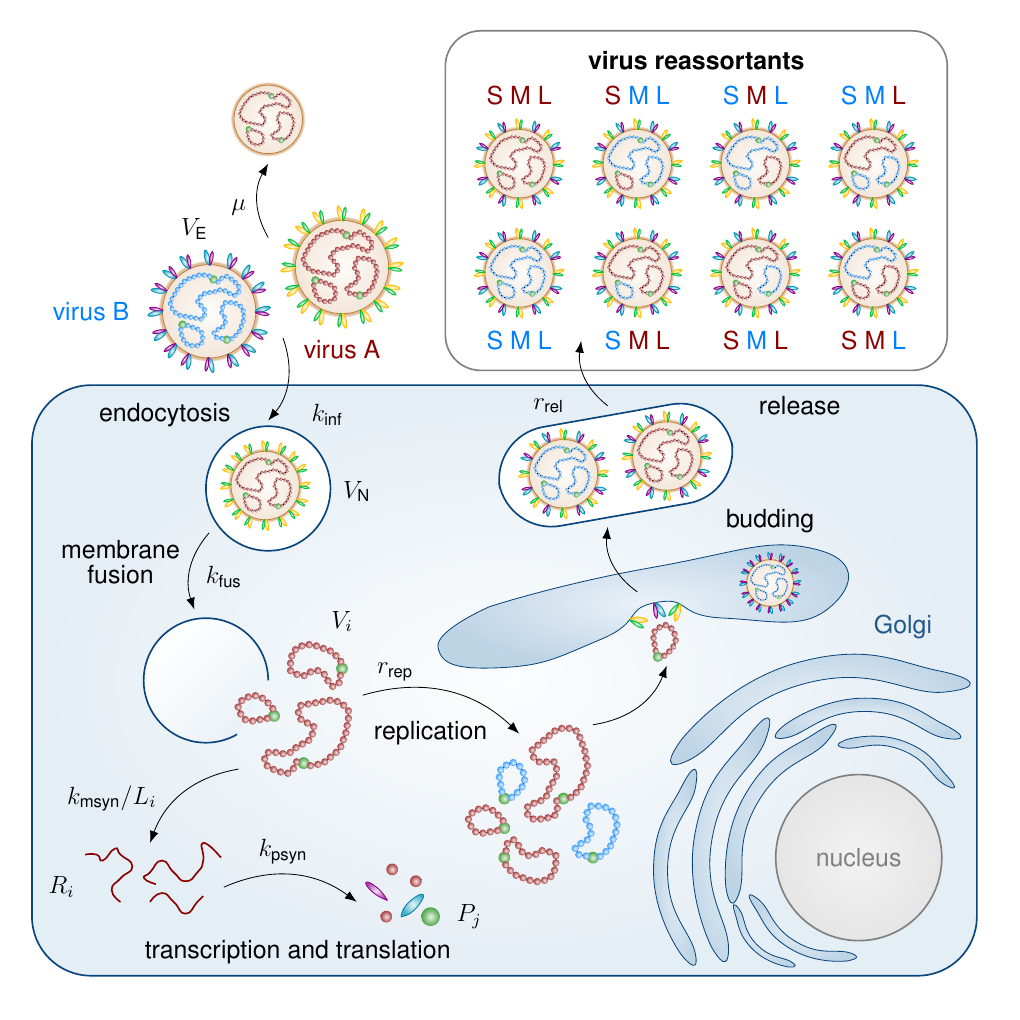} 
\end{tabular}
 \caption{A) If two viral strains, $V_A$ and $V_B$, are co-circulating, the target cells, $T$, of an
 infected host will become singly infected ($I_A$ and $I_B$), and  potentially  co-infected ($I_C$).
Co-infected cells have the potential to generate new viral progeny, different from that of
 the parental strains: $V_C \neq V_A$ and $V_C \neq V_B$.
  B) A co-infected cell can
 lead to reassortment events, and produce up to $2^3$ different reassortants. }
 \label{fig:reassortment}
 \end{figure}

Genomic reassortment is possible in viruses with segmented genomes, such as the Bunyaviruses, which themselves
include lethal pathogens of relevance to public health 
and of pandemic potential, {\em e.g.,} Lassa fever, Rift Valley fever and 
CCHF viruses~\cite{WHOblueprint2018}.
 Fig.~\ref{fig:reassortment}  illustrates the dynamics of reassortment at the cellular level for Bunyaviruses, or more generally for
 a tri-segmented virus. 
CCHFV is a tick-borne Bunyavirus,
with the potential to reassort, and an increasing geographical range due to the changing 
 climate~\cite{portillo2021epidemiological,fanelli2021risk}.
Understanding how adaptable to different hosts this potentially fatal human pathogen is, what role co-infection (as a first step to genomic reassortment) will play in the generation of potential new 
viral strains, and how those variants will spread among already infected ticks, is a challenge for theoretical biology.  

Due to the ability of ticks to carry multiple viruses or viral strains, epidemiologists have considered co-infection in tick-borne diseases~\cite{mosquera1998evolution,allen2005asymptotic,zhang2013co,slater2013modelling,yakob2013slaving,lass2013generating,maliyoni2019stochastic,cutler2021tick,pabon2023bayesian}, super-infection~\cite{mosquera1998evolution}, and co-transmission~\cite{gao2016coinfection,lou2017modelingco,vogels2019arbovirus}.
Specifically, epidemiologists are interested in quantifying (or understanding) the invasion potential of a novel virus or strain~\cite{Cross2005,meehan2020probability}, given the endemic ability of a resident one~\cite{alizon2013multiple}.
In the same way as $R_0$ provides conditions for successful establishment of a single virus in a susceptible population, the \emph{invasion reproduction number}, $R_I$, lends threshold conditions for successful invasion of a second virus when the population is
endemic with the first one~\cite{allen2019modelling,white2019dynamics}.
For instance, Gao {\em et al.} developed a
Susceptible-Infected-Susceptible (SIS) model for tick and host populations, and conducted a systematic analysis of invasion by the second virus~\cite{gao2016coinfection}.
More recently,
Bushman and Antia have developed
a general framework of the 
interaction between viral strains at the within-host level~\cite{bushman2019general}.
Pfab {\em et al.} have extended 
 the time-since-infection framework of Kermack and McKendrick~\cite{kermack1927contribution} for two pathogens~\cite{pfab2022time}.
 Rovenolt and Tate~\cite{rovenolt2022impact}
have developed a model of co-infection to study 
 how within-host interactions between parasites 
 can alter host competition
in an epidemic setting.
Thao Le {\em et al.}~\cite{le2022disentangling}
have studied a two-strain SIS model with
co-infection (or co-colonisation) which
incorporates 
 variation in transmissibility, duration of carriage, pair-wise susceptibility to co-infection, co-infection duration, and transmission priority effects. 
Finally, Saad-Roy {\em et al.}~\cite{saad2021superinfection}
have considered super-infection 
and its role during the 
the first stage of an infection on the evolutionary dynamics of the degree to which the host is asymptomatic.

In the case of plant pathogens, 
recent experimental studies have shown 
the complex nature of vector-virus-plant interactions 
and their role in the transmission and 
replication of viruses as single and co-infections~\cite{mclaughlin2022vector}.
Allen {\em et al.} modeled the transmission dynamics of viruses between vectors and plants, under the assumption that co-infection could only take place in plants~\cite{allen2019modelling}.
Chapwanya {\em et al.} developed
 a general deterministic epidemic model of crop-vector-borne disease for synergistic co-infection~\cite{chapwanya2021synergistic}.
Miller {\em et al.}
have shown that 
mathematical models on the kinetics of co-infection of 
plant
cells with two strains could not adequately describe the data~\cite{miller2022mathematical}.

Current mathematical models of co-infection need
to be put in perspective, as previously discussed by Lipsitch {\em et al.} and Alizon~\cite{lipsitch2009no,alizon2013co}.
Alizon compared different models of  co-infection and raised an issue of 
\emph{non-neutrality}~\cite{alizon2013co}.
He noticed that certain models of co-infection lead to an invasion reproduction number which does not tend to 
one, in the limit when the invasive and the resident pathogens are the same.
To solve this problem, Alizon proposed an alternative model structure, which includes a population of dually infected individuals with the resident pathogen, to achieve the desired neutral invasion reproduction number~\cite{alizon2013parasite,alizon2013co}.

In this paper, we first present a mathematical model of a single vector-borne virus to understand the role that different transmission routes play in the dynamics of the infected populations. Then, we study the dynamics of two different viruses, or viral strains, in a tick population making use of a classic co-infection model. After performing an invasion analysis, we explain the issue of
non-neutrality of the invasion reproduction number, and propose five  neutral alternatives. We conclude the paper with a summary of our alternative proposals, their applicability and limitations. 


\section{Mathematical model of a single viral strain in a population of ticks and their vertebrate hosts}
\label{sec:single-infection}

We consider a tick population feeding on a population of vertebrate hosts, where both populations are susceptible to  infection with virus $V_1$. 
The host and tick populations are divided into
 susceptible and infected subsets.
In what follows the number of susceptible hosts (ticks) is denoted $n_0$ ($m_0$), and the number of infected 
hosts (ticks) is denoted $n_1$ ($m_1$), respectively.

The mathematical model considers immigration,
death, viral transmission and recovery events in the populations;
namely, susceptible hosts and ticks immigrate into the population with rate $\Phi_H$ and $\Phi_T$, respectively.  
Susceptible and infected hosts can die with per capita rates $\mu_0$ and $\mu_1$, respectively, whereas susceptible and infected ticks are characterised by the per capita death rates $\nu_0$ and $\nu_1$, respectively. 
We assume an infected host can infect a susceptible tick with rate $\gamma_1$, and an infected tick can infect a susceptible host with rate $\beta_1$. 
Both of these transmission events involve a tick feeding on a vertebrate host, and are referred to as {\em systemic} transmission events~\cite{bhowmick2022ticks}.
The virus can additionally be transmitted from an infected tick to a susceptible one via co-feeding~\cite{gonzalez1992sexual,matser2009elasticity}. 
This occurs when ticks feed on a host in clusters, and close to each other; that is, on the same host and at the same time. In this instance the virus is transmitted by infected tick saliva, with this route of transmission referred to as {\em non-systemic}~\cite{bhowmick2022ticks}. 
We denote by $\alpha_1$ the rate at which an infected tick can infect a susceptible one via co-feeding.
We assume that  transmission events follow mass action kinetics.
For example, in the case of co-feeding, and with $m_0$ and $m_1$ the number of susceptible and infected ticks, respectively, the rate of infection for the susceptible population is $\alpha_1 m_0 m_1$.
Finally, once a tick contracts the virus, it remains infected  for life~\cite{gargili2017role}.
On the other hand, vertebrate hosts are characterised by short-lasting viremia~\cite{gargili2017role,spengler2019crimean}.
We, thus, assume that hosts clear the virus with rate $\varphi_1$~\cite{gonzalez1998biological,hoch2018dynamic}. %
The above set of events are brought together in the following system of ordinary differential equations (ODEs), which describe the dynamics of susceptible and infected hosts and ticks:
\begin{equation}
\label{eq:single}
    \begin{split}
        \frac{\dd n_0}{\dt} & = \Phi_H -\mu_0  n_0 - \beta_1 n_0 m_1 + \varphi_1 n_1, 
        \\
        \frac{\dd n_1}{\dt} & = -\mu_1 n_1 + \beta_1 n_0 m_1 - \varphi_1 n_1, 
        \\
        \frac{\dd m_0}{\dt} & = \Phi_T -\nu_0  m_0 - \gamma_1 m_0 n_1 - \alpha_1 m_0 m_1, 
        \\
        \frac{\dd m_1}{\dt} & = -\nu_1  m_1 + \gamma_1 m_0 n_1 + \alpha_1 m_0 m_1. 
    \end{split}
\end{equation}
We note that this system of ODEs~\eqref{eq:single} has a virus-free equilibrium (VFE), $(n_0^\star, 0, m_0^\star, 0)$, given by
\begin{equation}
\label{eq:df_equilibrium}
    n_0^\star=\frac{\Phi_H}{\mu_0}, \quad m_0^\star=\frac{\Phi_T}{\nu_0}.
\end{equation}

\subsection{Basic reproduction number}
\label{sec:single-basic}

The basic reproduction number, $R_0$, measures the mean number of new infections produced by an infected individual  (during its lifetime) in a population at the virus-free equilibrium; that is, when the population is completely susceptible~\cite{van2002reproduction}. 
$R_0$ (for the mathematical model~\eqref{eq:single}) can be calculated making use of the
 next-generation matrix method~\cite{diekmann1990definition}
 as follows.
 The 
 sub-system of differential equations for $(n_1, m_1)$
 is linearised  at the VFE, and its Jacobian, 
 $J$,
 is then written as $J\equiv T+V$, with $T$ the $2 \times 2$ matrix of transmission events which accounts for new infections in the susceptible population, and $V \equiv J-T$,
the $2 \times 2$ matrix
 tracking
the changes in the state of the infected populations~\cite{diekmann1990definition}.
The next-generation matrix is defined as $\mathbb{K} \equiv T (-V)^{-1}$, and the basic reproduction number, $R_0$, is given by the largest  eigenvalue of 
$\mathbb{K}$~\cite{diekmann1990definition}. For our system
we have
\begin{eqnarray}
J \equiv
\begin{pmatrix}
- \mu_1- \varphi_1 &   \beta_1 n_0^\star
\\
\gamma_1 m_0^\star  &  -\nu_1  +\alpha_1 m_0^\star
\end{pmatrix}
\; , 
\end{eqnarray}
 with
\begin{eqnarray}
T \equiv
\begin{pmatrix}
0 &   \beta_1 n_0^\star
\\
\gamma_1 m_0^\star  &  \alpha_1 m_0^\star
\end{pmatrix}
\; ,
\quad
\text{and}
\quad
V \equiv
\begin{pmatrix}
- \mu_1- \varphi_1 &   0
\\
0 &  -\nu_1 
\end{pmatrix}
\; ,
\end{eqnarray}
so that 
\begin{eqnarray}
\label{eq:ngm_single_infection}
    \mathbb{K} \equiv
\begin{pmatrix}
0 &   \beta_1 n_0^\star/\nu_1
\\
\gamma_1 m_0^\star/(\mu_1+ \varphi_1)  &  \alpha_1 m_0^\star/\nu_1
\end{pmatrix}
\; ,
\end{eqnarray}
which in turn implies
\begin{eqnarray}
\label{eq:R0_all_routes}
R_0 \equiv \frac{1}{2} \left[ 
\; 
\frac{\alpha_1 m_0^\star}{\nu_1} + \sqrt{
\left( \frac{\alpha_1 m_0^\star}{\nu_1} \right)^2
+ 4 \frac{\beta_1 \gamma_1 m_0^\star n_0^\star}{\nu_1(\mu_1+ \varphi_1)}}
\; 
\right]
\; .   
\end{eqnarray}
If $R_0<1$, the VFE is stable, and if 
$R_0>1$, it is  unstable.
The basic reproduction number can be rewritten as 
\begin{equation}
\label{eq:R0_function_of_routes}
    R_0 = \frac{1}{2} \left( R_{TT} + \sqrt{R_{TT}^2 + 4 R_{TH} R_{HT}} \; \right),
\end{equation}
where we have introduced the following type reproduction
numbers~\cite{heesterbeek2007type}
\begin{equation*}
    R_{TT} = \alpha_1 \frac{\Phi_T}{\nu_0 \nu_1}, \quad
    R_{TH} = \beta_1 \frac{\Phi_H}{\mu_0 \nu_1}, \quad
    R_{HT} = \gamma_1 \frac{\Phi_T}{\nu_0 (\mu_1+\varphi_1)},
\end{equation*}
which represent the contribution of each route of transmission, tick-to-tick, tick-to-host and host-to-tick, respectively,
to the total number of new
 infections  (of ticks and hosts)
in the susceptible population.  $R_{TT}$, $R_{TH}$, and $R_{HT}$ correspond to the entries of the next-generation matrix $\mathbb{K}$ (see Eq.~\eqref{eq:ngm_single_infection}). 
 $R_{HH} = 0$, since the virus cannot be directly transmitted from an infected host to a susceptible one. 
The expression of the basic reproduction number for a single virus (see Eq.~\eqref{eq:R0_function_of_routes}) clearly shows that co-feeding represents a singular route of  transmission,
compared to  systemic routes.
For example, $\beta_1$ (or $\gamma_1$) can
be very large, but if $\gamma_1$ (or $\beta_1$)
is negligible, the contribution to $R_0$
of viral systemic transmission will be negligible.
Therefore, co-feeding events (as
characterised by the parameter $\alpha_1$),
might maintain an epidemic if $R_{TT}>1$.
On the other hand,  systemic transmission requires both  tick-to-host and host-to-tick transmission routes to
be non-vanishing, so that
there is a chance for $R_0>1$, since as soon
as either $\beta_1$ or $\gamma_1$ are equal to zero,
$R_0=0$ in the absence of co-feeding.

We conclude this section mentioning a novel
 network approach (developed in Ref.~\cite{johnstone2020incorporating}),
 to compute the parameters
 $\alpha_1$,  $\beta_1$, and $\gamma_1$ 
 from first principles.
 It is reassuring to note that
 this approach leads to 
 a next-generation matrix
 with the same structure as 
$\mathbb{K}$ in
Eq.~\eqref{eq:ngm_single_infection}.

\subsection{Parameter values}
\label{sec:single-param}

We make use of recent literature to obtain parameter values for the ODE system~\eqref{eq:single}. 
Table~\ref{tab:pars_single_infection} contains a description of each model parameter, together with its plausible ranges and units.  Since infection with CCHFV~\cite{sorvillo2020towards,spengler2019crimean} or {\em Borrelia}~\cite{johnstone2020incorporating} is asymptomatic in ticks and vertebrate hosts (but unfortunately not in humans), we assume it does not affect their death rates; that is, $\mu_0=\mu_1\equiv \mu$ and $\nu_0=\nu_1\equiv\nu$~\cite{gao2016coinfection}. Given the narrow ranges in Table~\ref{tab:pars_single_infection} for $\Phi_H$, $\Phi_T$, $\varphi_1$ and $\mu$, we fix these parameters as follows: $\Phi_H = 1$ host per day, $\Phi_T = 2$ ticks per day, $\varphi_1 = 1/6$ per day, and $\mu = 10^{-3}$ per day. We derive plausible ranges for the other model parameters making use of Ref.~\cite{johnstone2020incorporating}, as illustrated in detail  in~\ref{sec:appendix-params}.

\begin{table}
    \centering
\resizebox{\textwidth}{!}{
\begin{tabular}{|c||c|c|c|c|}
\hline
     Parameter & Event & Range & Units & Reference \\
     \hline
        \hline
     $\beta_1$ & $H_0 + T_1 \to H_1 + T_1$ & $[10^{-7}, 10^{-5}]$ & 1/day/tick & \cite{johnstone2020incorporating} \\
     $\gamma_1$ & $T_0 + H_1 \to T_1 + H_1$ & 
     $[10^{-5}, 10^{-2}]$ & 1/day/host & \cite{johnstone2020incorporating} \\
     $\alpha_1$ & $T_0 + T_1 \to T_1 + T_1$ & $[10^{-6}, 10^{-4}]$ & 1/day/tick & \cite{johnstone2020incorporating} \\
     $\nu_0$ & Death rate of $T_0$ & $10^{-2}$ & 1/day & \cite{johnstone2020incorporating} \\
     $\nu_1$ & Death rate of $T_1$ & $10^{-2}$ & 1/day & \cite{johnstone2020incorporating} \\
     $\mu_0$ & Death rate of $H_0$ & $[2.8 \times 10^{-4}, 2.8 \times 10^{-3}]$ & 1/day & \cite{mpeshe2011mathematical} \\
     $\mu_1$ & Death rate of $H_1$ & $[2.8 \times 10^{-4}, 2.8 \times 10^{-3}]$ & 1/day & \cite{mpeshe2011mathematical} \\
     $\Phi_T$ & Arrival of ticks & [0.5, 3.5] & tick/day & \cite{bhowmick2022ticks} \\
     $\Phi_H$ & Arrival of hosts & [0.5, 1.5] & host/day & \cite{bhowmick2022ticks} \\
     $\varphi_1$ & $H_1 \to H_0$ & [1/7,1/5] & 1/day & \cite{hoch2018dynamic} \\
   \hline
\hline   
    $\alpha_2$ & $T_0 + T_2 \to T_2 + T_2$ & $[10^{-6}, 10^{-4}]$ & 1/day/tick & \cite{johnstone2020incorporating} \\
    $\delta_1$ & Transmission of $V_1$ by $T_c$ & $[10^{-6}, 10^{-4}]$ & 1/day/tick & Assumed \\
    $\delta_2$ & Transmission of $V_2$ by $T_c$ & $[10^{-6}, 10^{-4}]$ & 1/day/tick & Assumed \\
    $\kappa_1$ & Transmission of one copy of $V_1$ from $M_1$ & $[10^{-6}, 10^{-4}]$ & 1/day/tick & Assumed \\
    $\kappa_2$ & Transmission of one copy of $V_2$ from $M_2$ & $[10^{-6}, 10^{-4}]$ & 1/day/tick & Assumed \\
    $\epsilon_c$ & Probability of co-transmission & $[0,1]$ & - & - \\
    $\epsilon_1$ & Probability of dual transmission  of $V_1$ & $[0,1]$ & - & - \\
    $\epsilon_2$ & Probability of  dual transmission  of $V_2$ & $[0,1]$ & - & - \\
    $\nu_2$ & Death rate of $T_2$ & $10^{-2}$ & 1/day & \cite{johnstone2020incorporating} \\
    $\nu_c$ & Death rate of $T_c$ & $10^{-2}$ & 1/day & \cite{johnstone2020incorporating} \\
\hline
\end{tabular}
}
\caption{Model parameters introduced in~\eqref{eq:single} (top half), 
and 
\eqref{eqn:ticd_only_sys},
\eqref{eqn:within-host_invasion_sys},
\eqref{eq:alizon_model}, and
\eqref{eq:two-slot} 
(bottom half).}
\label{tab:pars_single_infection}
\end{table}

\subsection{Visualization of the basic reproduction number}
\label{sec:num-one}

\begin{figure}[h]
    \centering
    \includegraphics[width=\textwidth]{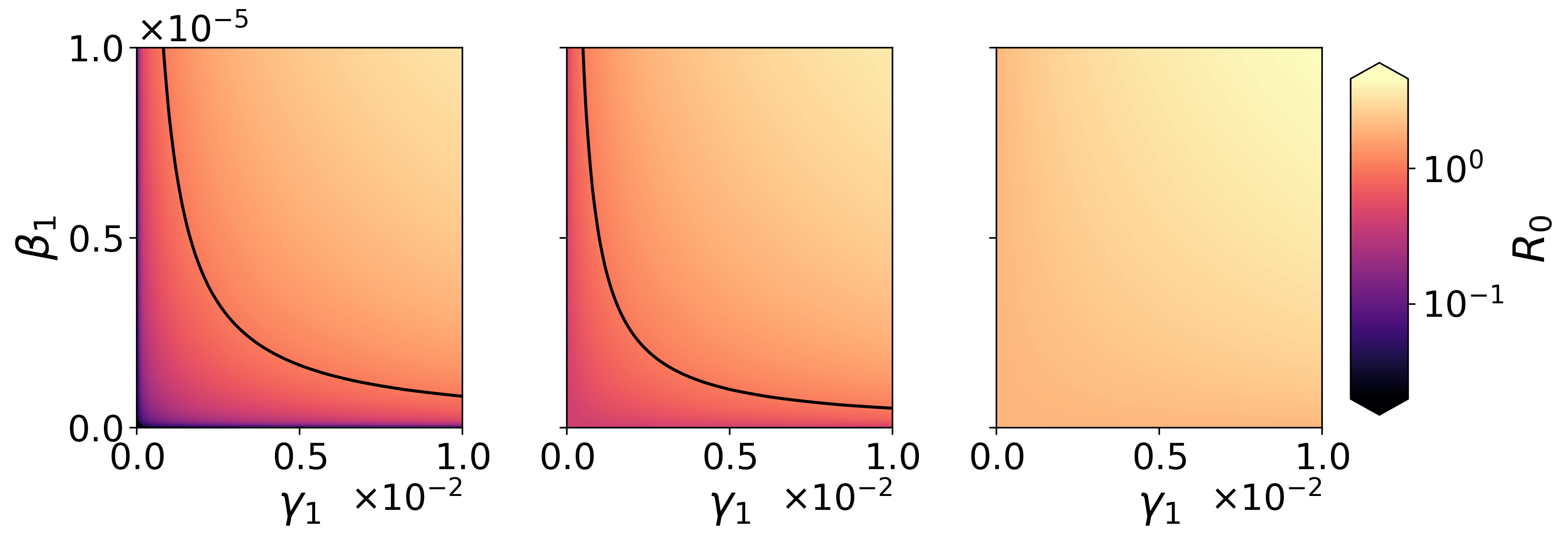}
    \caption{Contribution of $\alpha_1$, $\beta_1$ and $\gamma_1$ to the basic reproduction number, $R_0$, given by 
    Eq.~\eqref{eq:R0_all_routes}. Model parameters have been chosen as discussed in Section~\ref{sec:single-param} and units for $\alpha_1$, $\beta_1$ and $\gamma_1$ as in Table~\ref{tab:pars_single_infection}. On the left, $\alpha_1 = 10^{-6}$, in the middle $\alpha_1 = 2 \times 10^{-5}$, and on the right $\alpha_1 = 10^{-4}$. The parameters $\gamma_1$ and $\beta_1$ are varied along the $x$-axis and $y$-axis, respectively, from 0 to their maximum value listed in Table~\ref{tab:pars_single_infection}. Black curves represent the critical value $R_0 =1$.}
    \label{fig:SingleR0}
\end{figure}

We illustrate the dependence of $R_0$ on the transmission parameters $\alpha_1$, $\beta_1$ and $\gamma_1$, 
Fig.~\ref{fig:SingleR0}, making  use of~\eqref{eq:R0_all_routes},
and the parameter values from Section~\ref{sec:single-param}. 
Lighter colours correspond to greater values of $R_0$ (scale on right).
Black lines represent  a basic reproduction number equal to its critical value of 1.
We set $\Phi_H$, $\Phi_T$, $\varphi_1$, and $\mu$ to the values specified in Section~\ref{sec:single-param}, and set $\nu = 10^{-2}$ per day (see~\ref{sec:appendix-params}). 
 We consider a different value of $\alpha_1$ in each panel: on the left, $\alpha_1 = 10^{-6}$, in the middle $\alpha_1 = 2 \times 10^{-5}$, and on the right $\alpha_1 = 10^{-4}$ (units as provided in 
 Table~\ref{tab:pars_single_infection}). The corresponding values of $R_{TT}$ are $R_{TT} = 1.2 \times 10^{-2}$, $R_{TT} = 0.4$, and $R_{TT} = 2$. 
 Along the $x$-axis and $y$-axis we vary $\gamma_1$ and $\beta_1$, respectively, from $0$ to their maximum value listed in Table~\ref{tab:pars_single_infection}. We note that the area under the curve $R_0=1$ becomes smaller as $\alpha_1$ increases (from left to right), until it 
becomes zero when co-feeding transmission contributes to make $R_0$ greater than one on its own. As one would expect, smaller values of the transmission parameters $\alpha_1$, $\beta_1$ and $\gamma_1$ correspond to lower values of $R_0$ (purple regions on the left and middle panels).
Finally, we also note the symmetric role of
$\beta_1$ and $\gamma_1$ in $R_0$, as shown
in~\eqref{eq:R0_all_routes}.

\section{Two viral strains: tick population and co-feeding transmission} 
\label{sec:two-strain-infection}

In the previous section, we have shown that co-feeding can sustain an infection among ticks without systemic transmission.
The remainder of the paper will focus on the co-feeding route of transmission. 
We now move to the more complex case where multiple viral strains co-exist, introducing the notions of 
co-infection and co-transmission.
The population of ticks can be infected by two different circulating viral strains,
$V_1$ and $V_2$. 
$V_1$ is considered to be the resident strain and $V_2$ the invasive one ({\em e.g.}, one that emerges once
the tick population is endemic with $V_1$). 
The population of ticks can be classified by its infection status in four different compartments, as susceptible and infected ticks with the resident strain, $m_0$ and $m_1$, respectively, and
infected ticks with the invasive strain and co-infected
({\em i.e.,} infected with both strains $V_1$ and $V_2$) ticks, $m_2$ and $m_c$, respectively.
\begin{figure}[htp!]
\begin{subfigure}[b]{.45\textwidth}
\centering
\begin{tikzpicture}

    \node at (0,0) [circle,draw=black,align=center]  (a1) {$m_0$};
    \node at  (1.5,1.5) [circle,draw=black,align=center] (a2) {$m_1$};
    \node at (1.5,-1.5)  [circle,draw=black,align=center] (a3) {$m_2$};
    \node at  (3,0)  [circle,draw=black,align=center] (a6) {$m_c$};

    \draw[->,draw= cyan,thick] (a1)  -- node [pos=0.3,above=2.5pt]{\color{cyan}{$\lambda_1$}} (a2);
    \draw[->, draw=cyan,thick] (a1) -- node[below,midway,pos=0.3]{\color{cyan}{$\lambda_2$}} (a3);
    \draw[->,draw= magenta,thick] (a2)  -- node [above=15pt,midway,pos=1.3]{\color{magenta}{$\lambda_2+\lambda_c$}} (a6);
    \draw[->,draw=magenta,thick] (a3) -- node[below,right=1pt,pos=0.3]{\color{magenta}{$\lambda_1+\lambda_c$}} (a6);
    \draw[->,draw=dark-purple,thick] (a1) --  node[above,midway]{\color{dark-purple}{$\lambda_c$}} (a6);
    
\end{tikzpicture}
\caption{}
\label{fig:cofeeding-cotransmission}
\end{subfigure}
     \hfill
\begin{subfigure}[b]{.45\textwidth}
\centering
\begin{tikzpicture}

    \node at (0,0) [circle,draw=black,align=center]  (a1) {$m_0$};
    \node at  (1.5,1.5) [circle,draw=black,align=center] (a2) {$m_1$};
    \node at (1.5,-1.5)  [circle,draw=black,align=center] (a3) {$m_2$};
    \node at  (3,0)  [circle,draw=black,align=center] (a6) {$m_c$};

    \draw[->,draw = cyan,thick] (a1)  -- node [pos=0.3,above=2.5pt]{\color{cyan}{$\lambda_1$}} (a2);
    \draw[->, draw=cyan,thick] (a1) -- node[below,midway,pos=0.3]{\color{cyan}{$\lambda_2$}} (a3);
    \draw[->,draw= magenta,thick] (a2)  -- node [above=25pt,midway,pos=1.9]{\color{magenta}{$\ \phi_{2\mid 1}(\lambda_2+\lambda_c$)}} (a6);
    \draw[->,draw=magenta,thick] (a3) -- node[below,right=1pt,pos=0.3]{\color{magenta}{$\phi_{1 \mid 2}(\lambda_1+\lambda_c)$}} (a6);
    \draw[->,draw=dark-purple,thick] (a1) --  node[above,midway]{\color{dark-purple}{$\lambda_c$}} (a6);
\end{tikzpicture}
\caption{}
\label{fig:cofeeding-cotransmission-within_host}
\end{subfigure}
\hfill 
\begin{subfigure}[b]{0.45\textwidth}
\centering
\begin{tikzpicture}

    \node at (0,0) [circle,draw=black,align=center] (a1) {$m_0$};
    \node at (1.5,1.5) [circle,draw=black,align=center] (a2) {$m_1$};
    \node at (1.5,-1.5) [circle,draw=black,align=center] (a3) {$m_2$};
    \node at (3,3) [circle,draw=black,align=center] (a4) {$M_1$};
    \node at (3,-3) [circle,draw=black,align=center] (a5) {$M_2$};
    \node at (3,0) [circle,draw=black,align=center] (a6) {$m_c$};

    \draw[->,thick,cyan] (a1) -- node [pos=0.3,above=2.5pt]{\color{cyan}{$\lambda_1$}} (a2);
    \draw[->,thick,cyan] (a1) -- node [pos=0.3,below=2.0pt]{\color{cyan}{$\lambda_2$}} (a3);
    \draw[->,thick,cyan] (a2) -- node [pos=0.3,below=10pt,right=2pt]{\color{cyan}{$\lambda_1 + \lambda_{1,c} + \Lambda_1$}} (a4);
    \draw[->,thick,magenta] (a2) -- node [pos=0.3,above=10pt,right=2pt]{\color{magenta}{$\lambda_2 + \lambda_{2,c} + \Lambda_2$}} (a6);
    \draw[->,thick,cyan] (a3) -- node [pos=0.3,right=2pt]{\color{cyan}{$\lambda_2 + \lambda_{2,c} + \Lambda_2$}} (a5);
    \draw[->,thick,magenta] (a3) -- node [pos=0.3,above=10pt,right=2pt]{\color{magenta}{$\lambda_1 + \lambda_{1,c} + \Lambda_1$}} (a6);
    \draw[->,dark-purple,thick] (a1) to [out=90,in=180] node [pos=0.5, above] {$\Lambda_1 \quad$} (a4);
    \draw [dark-purple,thick,->] (a1) to [out=270,in=180] node [pos=0.5, below] {$\Lambda_2$}(a5);
    \draw[->,draw=dark-purple,thick] (a1) --  node[above,midway]{\color{dark-purple}{$\lambda_c$}} (a6);

\end{tikzpicture}
\caption{}
\label{fig:alizon}
\end{subfigure}
\hfill 
\begin{subfigure}[b]{0.45\textwidth}
\centering
\begin{tikzpicture}

    \node at (0,0) [circle,draw=black,align=center] (a1) {$m_0$};
    \node at (1.5,1.5) [circle,draw=black,align=center] (a2) {$m_1$};
    \node at (1.5,-1.5) [circle,draw=black,align=center] (a3) {$m_2$};
    \node at (3,3) [circle,draw=black,align=center] (a4) {$M_1$};
    \node at (3,-3) [circle,draw=black,align=center] (a5) {$M_2$};
    \node at (3,0) [circle,draw=black,align=center] (a6) {$m_c$};

    \draw[->,thick,cyan] (a1) -- node [pos=0.3,above=2.5pt]{\color{cyan}{$\lambda_1$}} (a2);
    \draw[->,thick,cyan] (a1) -- node [pos=0.3,below=2.0pt]{\color{cyan}{$\lambda_2$}} (a3);
    \draw[->,thick,cyan] (a2) -- node [pos=0.3,below=10pt,right=2pt]{\color{cyan}{$\lambda_1 + \lambda_{1,c} + \Lambda_1$}} (a4);
    \draw[->,thick,magenta] (a2) -- node [pos=0.3,above=10pt,right=2pt]{\color{magenta}{$\lambda_2 + \lambda_{2,c} + \Lambda_2$}} (a6);
    \draw[->,thick,cyan] (a3) -- node [pos=0.3,right=2pt]{\color{cyan}{$\lambda_2 + \lambda_{2,c} + \Lambda_2$}} (a5);
    \draw[->,thick,magenta] (a3) -- node [pos=0.3,above=10pt,right=2pt]{\color{magenta}{$\lambda_1 + \lambda_{1,c} + \Lambda_1$}} (a6);
    \draw[->,dark-purple,thick] (a1) to [out=90,in=180] node [left=1, above] {$\frac{\delta_1}{\delta_1 + \delta_2} \lambda_{1,c} + \Lambda_1$} (a4);
    \draw [dark-purple,thick,->] (a1) to [out=270,in=180] node [left=1, below] {$\frac{\delta_2}{\delta_1 + \delta_2} \lambda_{2,c} + \Lambda_2$}(a5);
    \draw[->,draw=dark-purple,thick] (a1) --  node[above,midway]{\color{dark-purple}{$\Lambda_c$}} (a6);

\end{tikzpicture}
\caption{}
\label{fig:two-slot}
\end{subfigure}
\caption{Illustrative diagrams of the mathematical models discussed in the paper for a population of co-feeding ticks with two viral strains. 
(a) Mathematical model of co-feeding transmission defined by Eq.~\eqref{eqn:ticd_only_sys}. Transmission
rates are defined in Eq.~\eqref{eq:pars_biased_model}. (b) Within-host mathematical model of co-feeding transmission defined by Eq.~\eqref{eqn:within-host_invasion_sys}.  Transmission
rates are defined in Eq.~\eqref{eq:pars_biased_model}. (c) Alizon's (generalised) proposal for co-infection and co-transmission described in Eq.~\eqref{eq:alizon_model}.  Transmission
rates are defined in Eq.~\eqref{eq:pars_Alizon_model}. (d) Two-slot mathematical model of co-infection and co-transmission defined in Eq.~\eqref{eq:two-slot}.  Transmission
rates are defined in Eq.~\eqref{eq:pars_Alizon_model} and Eq.~\eqref{eq:Lambda_c}.}
\end{figure}
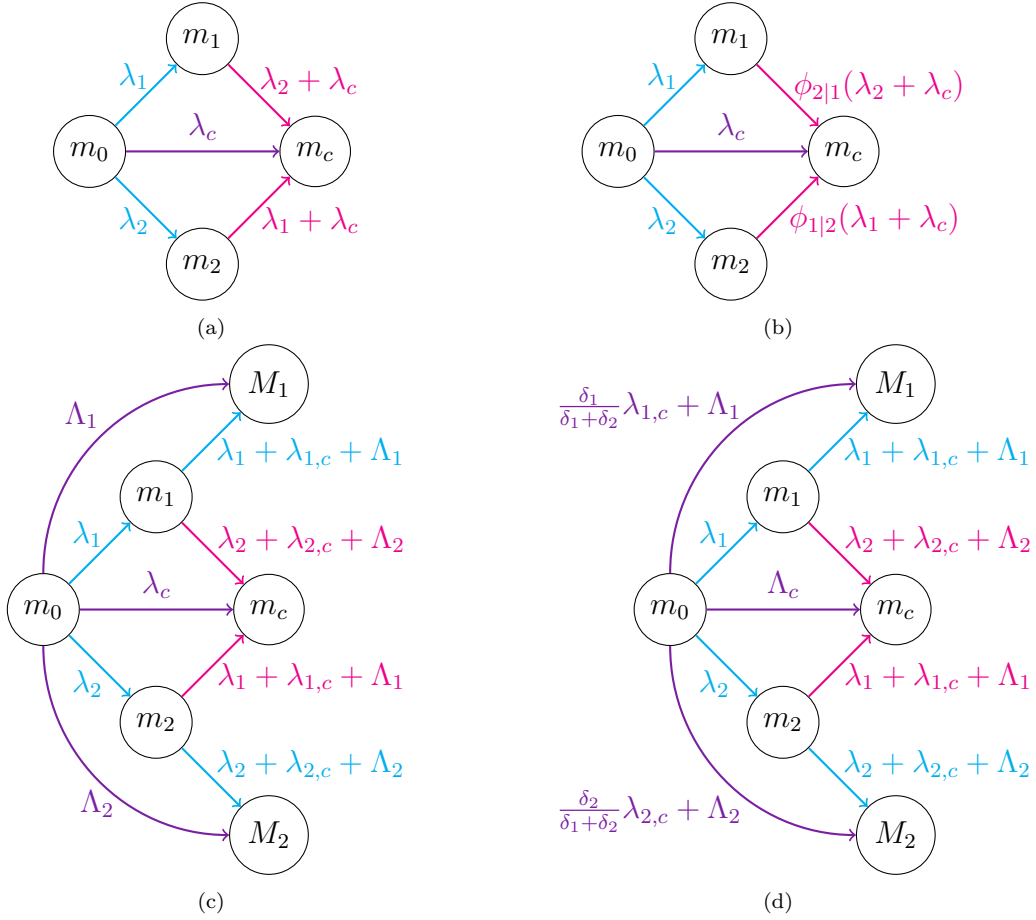
Figure~\ref{fig:cofeeding-cotransmission}
shows the mathematical model and the routes of viral transmission
considered
between  different tick compartments. The model corresponds to the
following system of ODEs:
\begin{equation} 
\label{eqn:ticd_only_sys}
\begin{split}
\frac{\dd m_0}{\dt}&= \Phi - \nu_0 m_0 - 
m_0 (\lambda_1 + \lambda_2 +\lambda_c)
\; ,
\\
\frac{\dd m_1}{\dt}&= - \nu_1 m_1 
+ m_0 \lambda_1
-m_1(\lambda_2 +\lambda_c)
\; ,
\\
\frac{\dd m_2}{\dt}&= - \nu_2 m_2 
+ m_0 \lambda_2  -m_2(\lambda_1 +\lambda_c)
\; ,
\\
\frac{\dd m_c}{\dt}&= - \nu_c m_c 
+ m_0\lambda_c
+m_1(\lambda_2 +\lambda_c)
+
m_2(\lambda_1 +\lambda_c)
\; ,
\end{split}
\end{equation}
where we have introduced
\begin{eqnarray}
\label{eq:pars_biased_model}
  \lambda_1&=&\alpha_1  m_1  + \delta_1 (1-\epsilon_c) m_c
  \; \nonumber,
  \\
  \lambda_2&=&\alpha_2  m_2 + \delta_2 (1-\epsilon_c) m_c
   \; ,
  \\
  \lambda_c &=& (\delta_1 + \delta_2) \epsilon_c m_c
  \; \nonumber,
\end{eqnarray}
with $\epsilon_c \in [0,1]$ representing the probability of co-transmission ($V_1$ and $V_2$).
We have assumed that 
the $m_1$  ($m_2$) population 
has  transmission parameter $\alpha_1$  ($\alpha_2$)
for $V_1$ ($V_2$),
and 
 the
 $m_c$ (co-infected) population 
has  transmission parameter $\delta_1$ 
for $V_1$ and 
$\delta_2$ 
for $V_2$, respectively.
We have also slightly abused notation by
writing $\Phi_T=\Phi$. 

\subsection{Basic reproduction number}
\label{sec:two-strain-basic}

The mathematical model defined by the system of ODEs~\eqref{eqn:ticd_only_sys} has a virus-free equilibrium (VFE), $(m_0^\star, 0, 0, 0)$, with $m_0^\star=\frac{\Phi}{\nu_0}$. To compute its basic reproduction number, we make use of the next-generation matrix method, as illustrated in detail in Section~\ref{sec:single-basic}. The $T$ and $V$ matrices are given by
\begin{eqnarray*}
T = \begin{pmatrix}
    \alpha_1 m_0^\star & 0 & \delta_1 (1-\epsilon_c) m_0^\star\\
    0 & \alpha_2 m_0^\star & \delta_2 (1-\epsilon_c) m_0^\star\\
    0 & 0 & (\delta_1 + \delta_2) \epsilon_c m_0^\star
\end{pmatrix}
\; ,
\quad
\text{and}
\quad
V = \begin{pmatrix}
    -\nu_1 & 0 & 0 \\
    0 & -\nu_2 & 0 \\
    0 & 0 & -\nu_c
\end{pmatrix}
\; .
\end{eqnarray*}
Thus, by computing the eigenvalues of the
next-generation matrix, ${\mathbb K} = T(-V)^{-1}$, the basic reproduction number of system~\eqref{eqn:ticd_only_sys} can be shown to be $R_0 = \max\{R_1, R_2, R_c\}$, with
\begin{equation*}
    R_1=\frac{\alpha_1}{\nu_1}m_0^\star, \quad R_2=\frac{\alpha_2}{\nu_2}m_0^\star, \quad R_c=\frac{(\delta_1+\delta_2)\epsilon_c}{\nu_c}m_0^\star.
\end{equation*}
Following the results from Ref.~\cite[Proposition 2.1]{gao2016coinfection}, we can explore the boundary equilibria of system~\eqref{eqn:ticd_only_sys}:

\begin{enumerate}

    \item The virus-free equilibrium, $E_0=(m_0^\star, 0, 0, 0)$, always exists.
    
    \item The endemic equilibrium with $V_1$, 
    $E_1=\left( \frac{\nu_1}{\alpha_1}, (1-\frac{1}{R_1})\frac{\Phi}{\nu_1}, 0, 0 \right) $, exists if and only if $R_1>1$.
    
    \item The endemic equilibrium with  $V_2$, $E_2=\left( \frac{\nu_2}{\alpha_2}, 0, (1-\frac{1}{R_2})\frac{\Phi}{\nu_2}, 0 \right)$, exists if and only if $R_2>1$.
    
    \item The endemic equilibrium with co-infected ticks, $E_c=\left( \frac{\nu_c}{(\delta_1+\delta_2)\epsilon_c}, 0, 0, (1-\frac{1}{R_c})\frac{\Phi}{\nu_c} \right)$, exists if and only if $\epsilon_c=1$, and $R_c>1$.
    
\end{enumerate}

\subsection{Invasion reproduction number}
\label{sec:two-strain-invasion}

We now assume that $R_1>1$, so 
that the endemic equilibrium $E_1$ of~\eqref{eqn:ticd_only_sys} exists. We write 
\begin{equation*}
    \overline{m}_0 = \frac{\nu_1}{\alpha_1}, \quad \overline{m}_1 =\left(1-\frac{1}{R_1}\right)\frac{\Phi}{\nu_1},
\end{equation*}
with $E_1 =(\overline{m}_0, \overline{m}_1, 0, 0)$. We aim to calculate the invasion reproduction number of $V_2$ by means of the 
next-generation matrix method. To this end, 
we identify  the
 invasive sub-system of $V_2$ of Eq.~\eqref{eqn:ticd_only_sys}, 
linearise it around  $E_1$,
compute its Jacobian matrix, and 
 define the $2 \times 2$ matrices $T$ and $V$. We can write
\begin{eqnarray*}
T \equiv \begin{pmatrix}
    \alpha_2 \overline{m}_0 & \delta_2 (1-\epsilon_c) \overline{m}_0 \\ \addlinespace
    \alpha_2 \overline{m}_1 & \substack{(\delta_1 + \delta_2) \epsilon_c (\overline{m}_0+\overline{m}_1) \\+ \delta_2 (1-\epsilon_c) \overline{m}_1}
\end{pmatrix}
\; ,
\quad
\,\text{and}
\quad \,
V \equiv \begin{pmatrix}
    -\alpha_1 \overline{m}_1 - \nu_2 & 0 \\
    \alpha_1 \overline{m}_1 & -\nu_c
\end{pmatrix}.
\end{eqnarray*}
The next-generation matrix, $\mathbb{K}=T(-V)^{-1}$, is given by
\begin{eqnarray*}
\mathbb{K} \equiv \begin{pmatrix}
    R_{22} & R_{c2} 
    \\
    R_{2c} & R_{cc}
\end{pmatrix}
\; ,
\end{eqnarray*}
with the type reproduction numbers $R_{22}$, $R_{2c}$, $R_{c2}$, and $R_{cc}$ given by
\begin{eqnarray*}
    R_{22} &=& \frac{\alpha_2\,\overline{m}_0}{\alpha_1\,\overline{m}_1 + \nu_2} + \frac{\alpha_1\,\overline{m}_1}{\alpha_1\,\overline{m}_1 + \nu_2}\,\frac{\delta_2\,(1-\epsilon_c)\,\overline{m}_0}{\nu_c}, 
    \\
    R_{c2} &=& \frac{\delta_2\,(1-\epsilon_c)\,\overline{m}_0}{\nu_c},
    \\
    R_{2c} &=& \frac{\alpha_2\, \overline{m}_1}{\alpha_1\,\overline{m}_1 + \nu_2} + \frac{\alpha_1\,\overline{m}_1}{\alpha_1\,\overline{m}_1 + \nu_2} \, \frac{(\delta_1+\delta_2)\, \epsilon_c\,(\overline{m}_0 + \overline{m}_1) + \delta_2\,(1-\epsilon_c)\,\overline{m}_1}{\nu_c}, 
    \\
    R_{cc} &=& \frac{(\delta_1+\delta_2)\, \epsilon_c\,(\overline{m}_0 + \overline{m}_1) + \delta_2\,(1-\epsilon_c)\,\overline{m}_1}{\nu_c}.
\end{eqnarray*}
The 
eigenvalues of $\mathbb{K}$ are solutions of  the 
following quadratic equation 
\begin{equation*}
    \lambda^2 - (R_{22} + R_{cc}) \lambda + (R_{22}\,R_{cc}-R_{c2}\,R_{2c})=0
    \; .
\end{equation*}
The invasion reproduction number, $R_I$, is the largest eigenvalue of $\mathbb{K}$, {\em i.e.,} 
\begin{equation}
\label{eq:RI}
    R_I = \frac{(R_{22}+R_{cc}) + \sqrt{(R_{22}+R_{cc})^2-4\,(R_{22}\,R_{cc}-R_{c2}\,R_{2c})}}{2}.
\end{equation}
When $R_I > 1$, that is, $R_{22} + R_{cc} - R_{22}\,R_{cc} + R_{c2}\,R_{2c} > 1$, $V_2$ is able to invade a tick population where the resident strain $V_1$ is endemic.


\section{Alternative neutral models of co-feeding, co-infection, and co-transmission}
\label{sec:problem}

The invasion reproduction number of the mathematical model from Section~\ref{sec:two-strain-invasion} is not {\em neutral}~\cite{alizon2013multiple,alizon2013parasite}.
By neutrality, we mean the following:
in the limit when the invasive strain tends to the resident
one, there should be no advantage for either strain, and thus,
$R_I \rightarrow 1$.
One can show for $R_I$ given
by Eq.~\eqref{eq:RI}  that
$R_I \centernot\rightarrow 1$.
In fact, 
we have
$R_I \rightarrow 1$ iff 
$\delta_2 = \frac{\alpha_1}{1+R_1}$,
and
$R_I \rightarrow 1$ iff 
$\delta_2 + \delta_1= \frac{\alpha_1}{R_1} $,
for 
$\epsilon_c=0$
and $\epsilon_c=1$, respectively, under the assumption that infection does not affect the death rate, {\em i.e.,} $\nu_0 = \nu_1 = \nu_2 = \nu_c$.
The issue of neutrality in co-infection models was brought up by Samuel Alizon in Ref.~\cite{alizon2013parasite} and Lipsitch {\em et al.} in Ref.~\cite{lipsitch2009no}.
We now present five alternative {\em neutral} formulations of the previous model. 
The first (and less optimal) option for obtaining a neutral model 
is to force  $R_I \rightarrow 1$ and in turn, consider the constraints this condition imposes on some of the model parameters.
The second one, as proposed by us to Samuel Alizon in private communication, is to consider
a normalised invasion reproduction number;
that is, define $R^N_I = \frac{R_I}{\lim_{2\rightarrow 1} R_I}$,
where by ${\lim_{2\rightarrow 1}  R_I}$, we mean
the value of the invasion reproduction number in the
limit when the invasive strain tends to the resident one (see Section~\ref{sec:normalised}). The third one generalises the mathematical 
model~\eqref{eqn:ticd_only_sys} by introducing the idea of within-host probability of  invasion (see Section~\ref{sec:within-host-invasion}).
The fourth one, as proposed by Alizon~\cite{alizon2013parasite},
is to consider a more general class of models, with 
doubly infected individuals (see Section~\ref{sec:alizon}).
A final one that we propose in Section~\ref{sec:two-slot}, is a generalisation of the approach of 
Alizon~\cite{alizon2013parasite,alizon2013multiple},
which clearly articulates the issue
of co-transmission.

\subsection{A normalised invasion reproduction number}
\label{sec:normalised}

The invasion reproduction number given by Eq.~\eqref{eq:RI}
is not neutral. Let us then define a {\em normalised}
invasion reproduction number, $R^N_I$, as follows
\begin{eqnarray}\label{eq:normed_RI}
    R^N_I &=& \frac{R_I}{\lim_{2\rightarrow 1} R_I}
    \; ,
\end{eqnarray}
where 
$\lim_{2\rightarrow 1}$ means
$\nu_2 \rightarrow \nu_1$,
$\alpha_2 \rightarrow \alpha_1$,
and 
$\delta_2 \rightarrow \delta_1$ for our co-infection model
(see Eq.~\eqref{eqn:ticd_only_sys}).
So defined, it is clear
 that 
${\lim_{2\rightarrow 1} R_I^N=1}$,
which is the desired neutrality condition.
We note that the condition for
the invasive strain to have the potential
to become  established is
$R_I > 1$.
Now that we have introduced a normalised
invasion reproduction number,
this condition becomes 
$ R_I^N > ({\lim_{2\rightarrow 1} R_I})^{-1}$.

\subsection{A model with within-host invasion}
\label{sec:within-host-invasion}

The fitness advantage of the invasive strain in model~\eqref{eqn:ticd_only_sys} stems from the assumption that $V_2$ can infect susceptible ticks and infected ticks by the resident strain with the same rate. However, this may not be realistic. For instance, a small amount of transmitted (invasive) virus may be less likely to establish infection in a tick that already has a high resident viral load, compared to a fully susceptible tick. The probability of
within-host invasion will, thus, depend on the relative within-host fitnesses of the invasive and resident strains. Therefore, we can adapt the previous model (Eq.~\eqref{eqn:ticd_only_sys}) by introducing the parameter $\phi_{i\mid j}$, which is the probability that strain $i$ can establish co-infection in a tick already infected by strain $j$, given that there is transmission of strain $i$ via co-feeding. This is similar to the super-infection framework described by Alizon in Ref.~\cite{alizon2013co}. The model can then be described by the following system of ODEs:
\begin{equation} 
\label{eqn:within-host_invasion_sys}
\begin{split}
\frac{\dd m_0}{\dt}&= \Phi - \nu_0 m_0 - 
m_0 (\lambda_1 + \lambda_2 +\lambda_c)
\; ,
\\
\frac{\dd m_1}{\dt}&= - \nu_1 m_1 
+ m_0 \lambda_1
-m_1\phi_{2\mid 1}(\lambda_2 +\lambda_c)
\; ,
\\
\frac{\dd m_2}{\dt}&= - \nu_2 m_2 
+ m_0 \lambda_2  -m_2\phi_{1\mid 2}(\lambda_1 +\lambda_c)
\; ,
\\
\frac{\dd m_c}{\dt}&= - \nu_c m_c 
+ m_0\lambda_c
+ m_1\phi_{2\mid 1}(\lambda_2 +\lambda_c)
+ m_2\phi_{1\mid 2}(\lambda_1 +\lambda_c)
\;,
\end{split}
\end{equation}
where  $\lambda_1, \lambda_2$ and $\lambda_c$ 
are defined by Eq.~\eqref{eq:pars_biased_model}.
The transmission events of this model
are summarised in Fig.~\ref{fig:cofeeding-cotransmission-within_host}. In~\ref{app:within-host_invasion} we show that the invasion reproduction number  for this model
satisfies the desired neutrality condition.

\subsection{A generalisation of Alizon's proposal}
\label{sec:alizon}

The mathematical model proposed by Alizon
in Ref.~\cite{alizon2013parasite}
to obtain a neutral invasion reproduction number requires two additional populations (see Fig.~\ref{fig:alizon}), namely the populations of doubly infected ticks with either $V_1$ or $V_2$, denoted by $M_1$ and $M_2$, respectively. Thus, there are six different tick compartments: $m_0$, susceptible ticks, $m_1, m_2$,  ticks (singly) infected with either $V_1$ or $V_2$, $M_1, M_2$,  doubly infected ticks  with either $V_1$ or $V_2$, and $m_c$,  co-infected ticks with both $V_1$ and $V_2$.
We note
that the co-infection models
in Ref.~\cite{alizon2013multiple}
do not consider co-transmission,
but  it is discussed in Ref.~\cite{alizon2013parasite}.
Thus, in what follows, and to develop a mathematical model of co-infection
and co-transmission in co-feeding ticks,
we  explain in detail
 what happens when a co-infected tick transmits virus to a
 singly infected tick. If co-transmission  of both, resident and invasive, strains occurs, the singly infected tick can only acquire one new viral strain, since  the mathematical model does not accommodate  triply infected ticks. Therefore,  if co-transmission takes place, a singly infected tick will acquire $V_1$ with probability $\frac{\delta_1}{\delta_1 + \delta_2}$,
 or  $V_2$ with probability $\frac{\delta_2}{\delta_1 + \delta_2}$. Hence, the overall rate at which a co-infected tick transmits $V_1$ to a singly infected tick is
$
    (1-\epsilon_c)\delta_1 + (\delta_1+\delta_2)\epsilon_c\frac{\delta_1}{\delta_1 + \delta_2} = \delta_1
    \; ,
$
where the first term represents transmission of 
$V_1$ if no co-transmission, and the second
term represents 
transmission of 
$V_1$ in the event of  co-transmission.
Similarly, the  overall rate  a co-infected tick transmits $V_2$ to a singly infected tick is $\delta_2$.
We now write down the system of ODEs for Alizon's generalised mathematical model of a co-feeding tick population,
with two circulating viral strains, which allows  for co-infection
and co-tranmission, and at most doubly infected ticks
(with the same strain $M_1$ and $M_2$, or with different ones $m_c$).
We have
\begin{equation}
\label{eq:alizon_model}
\begin{split}
        \frac{\dd m_0}{\dd t} &= \Phi - \nu_0\,m_0 - m_0\,(\lambda_1 + \lambda_2 +  \lambda_{1,c} + \lambda_{2,c} + \Lambda_1 + \Lambda_2), 
        \\
        \frac{\dd m_1}{\dd t} &= -\nu_1\,m_1 + m_0\,\lambda_1 - m_1\,(\lambda_1 + \lambda_2 +  \lambda_{1,c} + \lambda_{2,c} + \Lambda_1 + \Lambda_2),
        \\
        \frac{\dd m_2}{\dd t} &= -\nu_2\,m_2 + m_0\,\lambda_2 - m_2\,(\lambda_1 + \lambda_2 +  \lambda_{1,c} + \lambda_{2,c} + \Lambda_1 + \Lambda_2),
        \\
        \frac{\dd m_c}{\dd t} &= -\nu_c\,m_c + m_0\,(\lambda_{1,c} + \lambda_{2,c}) + m_1\,(\lambda_2 + \lambda_{2,c} + \Lambda_2) + m_2\,(\lambda_1 + \lambda_{1,c} + \Lambda_1),
        \\
        \frac{\dd M_1}{\dd t} &= -\upsilon_1\,M_1 + m_0\,\Lambda_1 + m_1\,(\lambda_1 + \lambda_{1,c} + \Lambda_1) ,
        \\
        \frac{\dd M_2}{\dd t} &= -\upsilon_2\,M_2 + m_0\,\Lambda_2 + m_2\,(\lambda_2 + \lambda_{2,c} + \Lambda_2), 
\end{split}
\end{equation}
where we define
\begin{equation}
\label{eq:pars_Alizon_model}
\begin{split}
  \lambda_1 & = \alpha_1  m_1  + \delta_1 (1-\epsilon_c) m_c
  + 2 \kappa_1 (1-\epsilon_1) M_1, \\
  \lambda_2 & = \alpha_2  m_2 + \delta_2 (1-\epsilon_c) m_c
  + 2 \kappa_2 (1-\epsilon_2) M_2, \\
  \lambda_{1,c} & = \delta_1 \epsilon_c m_c, \\
  \lambda_{2,c} & = \delta_2 \epsilon_c m_c, \\
  \Lambda_1 & = 2\kappa_1 \epsilon_1 M_1, \\
  \Lambda_2 & = 2\kappa_2 \epsilon_2 M_2,
\end{split}
\end{equation}
with $\epsilon_c$,  the probability of co-transmission of the two viral strains by a co-infected ($m_c$) tick, and with $\epsilon_1$ ($\epsilon_2$) the probability of co-transmission by  a doubly 
infected tick $M_1$ ($M_2$), respectively. 
The original
model of co-infection with co-transmission 
defined in Ref.~\cite{alizon2013parasite} assumed
 $\epsilon_1 =\epsilon_2=\epsilon_c = \epsilon$.
 We warn the reader that we have defined $\lambda_1$ and $\lambda_2$ to mean two different things in 
 Eq.~\eqref{eq:pars_Alizon_model}
 and Eq.~\eqref{eq:pars_biased_model}. We shall always clarify
 in what follows, which of the two definitions is implied.
Fig.~\ref{fig:alizon} shows the transmission events described by Eq.~\eqref{eq:alizon_model}.
We note that co-transmission by the $M_1$ tick population 
to a susceptible tick implies
double transmission of the resident viral strain $V_1$,
and that 
co-transmission by the $M_2$ tick population
(to a susceptible tick) implies
double transmission of the resident strain $V_2$.
Finally, the parameter
$\kappa_1$ ($\kappa_2$) is the rate of
transmission  of a single copy of $V_1$ ($V_2$)
from an $M_1$ ($M_2$) tick to a susceptible
one; thus, the factor of $2$ in the previous expression for $\Lambda_1$  ($\Lambda_2$).
We remind the reader that the model defined above
includes death, immigration and transmission events.
We have assumed each tick compartment has a different
death rate, and  immigration replenishes the
susceptible tick compartment.
In~\ref{app:alizon} we carefully derive the invasion reproduction number of this model and show 
 its neutrality.

\subsection{Two-slot model of co-feeding, co-infection and co-transmission}
\label{sec:two-slot}

In the model defined by Eq.~\eqref{eq:alizon_model}, a co-transmission event from a co-infected tick, in the $m_c$ compartment, to a susceptible tick implies the transmission of both viral strains at once. Here we extend the 
previous model to allow for the possibility that such a co-transmission event could instead result in the transmission of two copies of $V_1$ or two copies of $V_2$. 
The idea of this generalised {\em two-slot model} is
as follows:
since ticks can be at most doubly infected,
we assume each tick has 
two {\em infection slots} that can be occupied (or not). In the previous model, ``co-transmission'' to a susceptible tick meant transmission of both viral strains.
In this model 
``co-transmission'' means occupying both slots,
in such a way, that the slots can be occupied by 
two copies of the same virus (leading to 
$M_1$ or $M_2$ ticks), or two different strains
(leading to $m_c$ ticks).
The dynamics of the two-slot model can be written as
\begin{equation}\label{eq:two-slot}
    \begin{split}
        \frac{\dd m_0}{\dt} & = \Phi - \nu_0 m_0 - 
        m_0 (\lambda_1 + \lambda_2 + \lambda_{1,c} + \lambda_{2,c} + \Lambda_1 + \Lambda_2),
        \\
        \frac{\dd m_1}{\dt} & = - \nu_1 m_1 
        + m_0 \lambda_1
        -m_1(\lambda_1 + \lambda_2 + \lambda_{1,c} + \lambda_{2,c} + \Lambda_1 + \Lambda_2),
        \\
        \frac{\dd m_2}{\dt} & = - \nu_2 m_2 
        + m_0 \lambda_2  -m_2(\lambda_1 + \lambda_2 + \lambda_{1,c} + \lambda_{2,c} + \Lambda_1 + \Lambda_2), 
        \\
        \frac{\dd m_c}{\dt} & = - \nu_c m_c 
        + m_0 \Lambda_c
        + m_1( \lambda_2 + \lambda_{2,c} + \Lambda_2)
        + m_2( \lambda_1 + \lambda_{1,c} + \Lambda_1), 
        \\
        \frac{\dd M_1}{\dt} & = - \upsilon_1 M_1 + m_0 \left(\Lambda_1 + \frac{\delta_1}{\delta_1 + \delta_2} \lambda_{1,c} \right)
        + m_1 (\lambda_1  + \lambda_{1,c} + \Lambda_1), 
        \\
        \frac{\dd M_2}{\dt} & = - \upsilon_2 M_2 + m_0 \left(\Lambda_2 + \frac{\delta_2}{\delta_1 + \delta_2} \lambda_{2,c} \right)
        + m_2 (\lambda_2  + \lambda_{2,c} + \Lambda_2),
    \end{split}
\end{equation}
where
$\lambda_1, \lambda_2, \lambda_{1,c}, 
\lambda_{1,c}, \Lambda_1$, and $\Lambda_2$
have been defined in
Eq.~\eqref{eq:alizon_model}, and
with  $\Lambda_c$ given by
\begin{equation}
\label{eq:Lambda_c}
    \Lambda_c = \displaystyle{\frac{2 \delta_1 \delta_2}{\delta_1 + \delta_2}} \epsilon_c m_c.
\end{equation}
In~\ref{app:two-slot} we 
 describe in great detail the transmission  events considered in 
the two-slot mathematical model, show the existence of an endemic equilibrium for $V_1$, and prove that the model
leads to a neutral invasion reproduction number.

\subsection{Numerical study of the invasion reproduction number}
\label{sec:numerical-invasion}

In Section~\ref{sec:two-strain-infection} we have defined
and computed the invasion reproduction number
for a mathematical model
of co-infection and co-transmission in co-feeding ticks.
We have argued that such a model is not neutral,
and have in turn proposed different mathematical
models which do not suffer from such problem.
We now propose a numerical study
of the invasion reproduction number for
the ``not-neutral'' model 
 introduced in Section~\ref{sec:two-strain-infection},
 as well as  the invasion reproduction number for the
 model solutions proposed above to guarantee neutrality.

 In what follows we assume that
$
     \kappa_1 = \sfrac{ \alpha_1}{2}, \quad
     \text{and} \quad
     \kappa_2 = \sfrac{ \alpha_2}{2},
$
which are appropriate choices when considering
viral infections or 
micro-parasites~\cite{alizon2013parasite}.
As discussed by Alizon in 
Ref.~\cite{alizon2013parasite},
 if an infected host  (by a certain viral strain)
 is re-infected by the exact same strain, we do not expect to see a change in its viral load (and hence in transmission rate). 
 For co-infected ticks in the $m_c$ compartment
 (and infected by both
 $V_1$ and $V_2$), it is reasonable
 to hypothesise that potential within-tick
 interactions between the two strains do not lead to a change in transmission rates, when
 compared to doubly infected ticks in the $M_1$
 or $M_2$ compartments.  Thus,
 we assume $\delta_1 = \kappa_1$ and $\delta_2 = \kappa_2$.
Finally, and as justified earlier, we  
 set $\nu_0 = \nu_1 = \nu_2 = \nu_c = \upsilon_1 = \upsilon_2 = 10^{-2}$ per day. We also fix 
the immigration rate to be $\Phi=2$ ticks per day, and 
$\alpha_1=10^{-4}$ per tick per day. 
These choices lead to a basic reproduction number of $R_1=2$ for the resident strain, $V_1$.

\begin{figure}
\includegraphics[width = \textwidth]{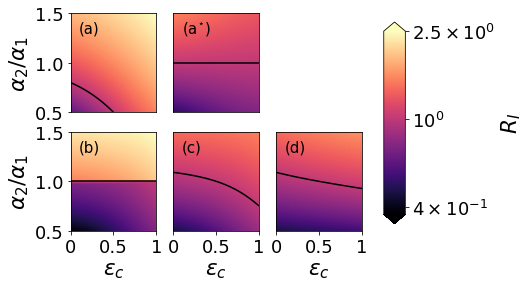}
\caption{Heatmaps of  the invasion reproduction number for (a) the ``not-neutral'' model~\eqref{eq:RI}, 
(a$^{\star}$) the normalised proposal~\eqref{eq:normed_RI}, (b) the  within-host model  with $\phi_{i\mid j}$, Eq.~\eqref{eqn:within-host_invasion_sys}, (c) Alizon's model with co-transmission~\eqref{eq:alizon_model}, and (d) the two-slot mathematical model~\eqref{eq:two-slot}. 
The $x$-axis represents $\epsilon_c$, the probability of co-transmission from  co-infected ticks.
The $y$-axis shows
the ratio $\alpha_2/\alpha_1$ in the range $[0.5,1.5]$.
We set
$\alpha_1 = 10^{-4}$ per tick per day.
 Black lines mark the contours where the invasion reproduction number, $R_I$, is equal to 1. For Alizon's model and the two-slot extension, we have set $\epsilon_1=\epsilon_2=\sfrac{1}{2}$.}
\label{fig:heatmaps_RI}
\end{figure}

In 
Fig.~\ref{fig:heatmaps_RI} we compare how different values of the transmission parameter, $\alpha_2$, and the co-transmission probability, $\epsilon_c$, affect the invasion reproduction number, $R_I$, computed in the ``not-neutral'' scenario~\eqref{eq:RI} (panel (a)), in the normalized proposal of Eq.~\eqref{eq:normed_RI} (panel (a$^{\star}$)), in the 
within-host model~\eqref{eqn:within-host_invasion_sys} (panel (b)), in Alizon's model with co-transmission~\eqref{eq:alizon_model} (panel (c)), and in the two-slot mathematical model~\eqref{eq:two-slot} (panel (d)). In particular, $\epsilon_c$ is varied along the $x$-axis, whereas the ratio $\sfrac{\alpha_2}{\alpha_1}$ is varied from $0.5$ to $1.5$ along the $y$-axis. Black lines mark the contours where the invasion reproduction number, $R_I$, is equal to 1. For Alizon's model and the two-slot extension, we have set $\epsilon_1=\epsilon_2=\sfrac{1}{2}$.
For the  within-host model, we define the probability that strain $i$ can establish co-infection in a tick already infected by strain $j$ as follows
\begin{equation*}
    \phi_{i\mid j} = \begin{cases}
    1,&  \text{if} \quad \alpha_i>\alpha_j,\\
    0,  &  \text{if} \quad  \alpha_i\leq \alpha_j.
    \end{cases}
\end{equation*}
We note that the highest values of the invasion reproduction number occur in panel (a) and (b) of Fig.~\ref{fig:heatmaps_RI}. In panel (a), the invasion reproduction number is clearly not neutral, since $R_I>1$ when $\alpha_2 = \alpha_1$, and also for some regions of parameter space with $\alpha_2<\alpha_1$. For this model, if  infected ticks with the invasive strain, $V_2$, are rare compared to ticks infected with the resident strain, $V_1$, then $V_2$ has an initial advantage over $V_1$. Each tick infected with the invasive strain  has the opportunity to infect a much larger number of ticks ($\overline{m}_0 + \overline{m}_1$), than 
those which can be infected by  a tick from the ${m}_1$ compartment. This allows $V_2$ to invade the $V_1$ endemic system, for large enough values of $\alpha_2$ and $\epsilon_c$. The co-transmission probability, $\epsilon_c$, affects the value of $R_I$, since it changes the rate at which co-infected ticks transmit $V_1$ and $V_2$ to susceptible ticks, $m_0$.  These rates are $\delta_1 +\epsilon_c\delta_2$ and $\delta_2 + \epsilon_c\delta_1$, respectively. Therefore a higher probability of co-transmission enables both strains to be transmitted more often. For the normalised invasion reproduction number in panel (a$^{\star}$), $R_I=1$ when $\alpha_2=\alpha_1$, for every value of $\epsilon_c$, given its definition. When $\alpha_2\neq \alpha_1$, $R_I$ does depend on $\epsilon_c$, but less so than   for the model of panel (a), since $\lim_{2\to 1} R_I$ increases with $\epsilon_c$. In panel (c), showing the invasion reproduction number for Alizon's model, when $\epsilon_c=\sfrac{1}{2}$ (equal to $\epsilon_1$ and $\epsilon_2$), $R_I=1$ for $\alpha_2 = \alpha_1$. As $\epsilon_c$ increases, so does the value of $R_I$, since a higher co-transmission probability enables $V_2$ to be transmitted along with $V_1$ more often. The invasion reproduction number of the two-slot model in panel (d) behaves in a similar fashion. However, increasing $\epsilon_c$ does not give as much of an advantage to the invasive strain, since co-transmission events can result in the transmission of two copies of $V_1$.

\section{Discussion and conclusions}

In this paper, we consider the role of  
different transmission routes for
a single vector-borne virus
in a population of ticks and vertebrate hosts.
We then study 
co-infection  and co-transmission 
of two circulating vector-borne viral strains
in a population of
co-feeding ticks.
We define and compute both the basic reproduction number and the invasion reproduction number, which provides the conditions under which a new variant can emerge (possibly endogenously from genomic reassortment).
We illustrate how a classic and {\em intuitive} model of invasion was not, in fact, neutral with respect to the invading strain;
that is, using this model to understand, for example, the minimum selective advantage that needs to be present for a invading strain to take hold of an endemic population (with the resident one) will privilege one strain over the other.
This is not a problem {\em per se}, as it might be the correct model from a mechanistic perspective.
However, it is important to characterise the
underlying properties of a mathematical model, especially if it is intended to be used as part of an inference procedure.
We also presented several alternative formulations of co-infection and co-transmission models that are, by definition neutral.
We have shown that each model has distinct and specific behaviour concerning the invasion reproduction number.
The take-home message of this review is that the assumptions used to model these important and complex infection systems matter,
specially 
when making inferences about pathogens
 of potential pandemic
emergence. 
In the real world, the choice
of model, from the different alternatives
presented and discussed here, will
clearly depend on the virus, as well
as the 
 immunology and ecology of the hosts those viruses infect. 

In conclusion, we note that while we have focused on deterministic models of tick-borne disease transmission, stochastic analogous may be considered instead, particularly when studying the invasion potential of a rare circulating viral strain~\cite{belluccini2023stochastic,maliyoni2019stochastic,maliyoni2023multipatch}. In a stochastic framework, the reproduction number is defined as a random variable rather than as an average~\cite{artalejo2013exact}, since its distribution encodes the probability of an epidemic occurring if a pathogen is introduced into a fully susceptible population by a small number of infected individuals~\cite{artalejo2013exact}. Thus, future work should include a study of the
 invasion reproduction number
probability distribution, 
as well as an exploration of 
 the issue of non-neutrality making
 use of stochastic approaches~\cite{yuan2011stochastic}.
Finally, given recent reports which
indicate an increase in the number of Zika and Dengue virus co-infection
cases  in expanding co-endemic regions~\cite{lin2023co}, 
it is of utmost importance to have suitable
within-host mathematical models  to
study the
impact of co-infection on viral infection dynamics.


\appendix

\section{Parameters from network approach}
\label{sec:appendix-params}

Since  CCHFV  infection is asymptomatic in ticks and vertebrate hosts, we assume that it does not affect their death rates; that is, $\mu_0=\mu_1\equiv \mu$ and $\nu_0=\nu_1\equiv\nu$.
Moreover, for the purposes of this section, we fix the values of some parameters within the ranges in Table~\ref{tab:pars_single_infection} as follows: $\Phi_H = 1$ host per day, $\Phi_T = 2$ ticks per day, $\varphi_1 = 1/6$ per day, and $\mu = 10^{-3}$ per day. We now derive plausible ranges for the other model parameters
making use of the results from Ref.~\cite{johnstone2020incorporating}.
From Ref.~\cite[Equation (19)]{johnstone2020incorporating}, we identify
\begin{equation*}
        R_{TT} = \sigma \nu_{ln} \langle k_{out} \rangle
        = \frac{\alpha_1 \Phi_T}{\nu_0 \nu_1}.      
\end{equation*}
From Ref.~\cite{johnstone2020incorporating}
$\langle k_{out} \rangle = \frac{\Phi_T}{\nu}$, with $\langle k_{out} \rangle = 2 \times 10^{2}$, from which 
we obtain $\nu = 10^{-2}$ per day, given the fixed value of
$\Phi_T$. 
Ref.~\cite{johnstone2020incorporating} also provides the following values:
$\sigma = 0.1$ and $\nu_{ln} = 2.4 \times 10^{-3}$ or $\nu_{ln} = 2 \times 10^{-2}$ depending on the disease~\cite[Table 2]{johnstone2020incorporating}.
Thus, we derive $\alpha_1 = \sigma  \nu_{ln} \nu$,
so that we conclude $\alpha_1 = 2.4 \times 10^{-6}$ or $\alpha_1 = 2 \times 10^{-5}$ per day per tick, respectively. From here, we assume values of $\alpha_1$ in the interval $\left[10^{-6},10^{-4}\right]$.

We now turn to the
 transmission parameters (tick-to-host) $\beta_1$ and (host-to-tick) $\gamma_1$.
 From Ref.~\cite[Equations (6) and (7)]{johnstone2020incorporating}, we have
\begin{equation*}
    R_{TH} = \sigma \nu_{hn} = \frac{\beta_1 \Phi_H}{\mu_0 \nu_1}, \quad R_{HT} = \nu_{lh} \langle k_{out} \rangle = \frac{\gamma_1 \Phi_T}{\nu_0 (\mu_1+\varphi_1)}.
\end{equation*}
Thus, we can write $\beta_1 = \sigma  \mu  \nu \nu_{hn} \sim 10^{-6}$, from which we define a plausible range for $\beta_1$ as $\left[10^{-7}, 10^{-5}\right]$. Finally, $\gamma_1 = \nu_{lh}  (\mu + \varphi_1)$. As the value of $\nu_{lh}$ depends on the disease (it is either $\nu_{lh} = 1.9 \times 10^{-3}$ or $\nu_{lh} = 7.3 \times 10^{-2}$), we consider the interval $\left[10^{-5}, 10^{-2}\right]$ for $\gamma_1$.

\section{Within-host invasion model}
\label{app:within-host_invasion}

Following the same steps as those provided in Section~\ref{sec:two-strain-invasion}, the next-generation matrix  for the within-host invasion model is given  by
\begin{eqnarray*}
\mathbb{K} \equiv \begin{pmatrix}
    b_{11} & b_{12} \\
    b_{21} & b_{22}
\end{pmatrix}
\; ,
\end{eqnarray*}
where
\begin{eqnarray*}
    b_{11} &=& \frac{\alpha_2\,\overline{m}_0}{\alpha_1\,\phi_{1\mid 2}\overline{m}_1 + \nu_2} + \frac{\alpha_1\,\phi_{1\mid 2}\overline{m}_1}{\alpha_1\,\phi_{1\mid 2}\overline{m}_1 + \nu_2}\,\frac{\delta_2\,(1-\epsilon_c)\,\overline{m}_0}{\nu_c}, \\
    b_{12} &=& \frac{\delta_2\,(1-\epsilon_c)\,\overline{m}_0}{\nu_c}, \\
    b_{21} &=& \frac{\alpha_2\,\phi_{2\mid 1} \overline{m}_1}{\alpha_1\,\phi_{1\mid 2}\overline{m}_1 + \nu_2} + \frac{\alpha_1\,\phi_{1\mid 2}\overline{m}_1}{\alpha_1\,\phi_{1\mid 2}\overline{m}_1 + \nu_2} \, \frac{(\delta_1+\delta_2)\, \epsilon_c\,(\overline{m}_0 + \phi_{2\mid 1}\overline{m}_1) + \delta_2\,(1-\epsilon_c)\,\phi_{2\mid 1}\overline{m}_1}{\nu_c}, \\
    b_{22} &=& \frac{(\delta_1+\delta_2)\, \epsilon_c\,(\overline{m}_0 + \phi_{2\mid 1}\overline{m}_1) + \delta_2\,(1-\epsilon_c)\,\phi_{2\mid 1}\overline{m}_1}{\nu_c}.
\end{eqnarray*}
When considering neutrality (\emph{i.e.,} the invasive strain is the same as the resident strain), all infected tick populations ($m_1$, $m_2$, and $m_c$) are infected with the same  viral strain. Thus, we have $\nu_2 = \nu_c = \nu_1$ and $\alpha_2 = \alpha_1$. We also set $\delta_1 = \delta_2 = \frac{\alpha_1}{2}$, representing that ticks in the $m_c$ compartment will transmit virus at the same overall rate as ticks in the $m_1$ compartment ({\em i.e.,} $\delta_1 + \delta_2 = \alpha_1$). Furthermore, we would expect $\phi_{1\mid 2} = \phi_{2\mid 1} = 0$, since in the within-host environment (a tick) the transmitted strain is likely to be rare compared to the established strain and if the resident strain is the same as the invasive strain, then both strains have the same within-host fitness, implying that the rare transmitted strain will have no within-host advantage over the established strain, and will be unable to establish  co-infection in the host (tick). With these limits, the elements of the next-generation matrix become
\begin{eqnarray*}
    b_{11} &=& \frac{\alpha_1\,\overline{m}_0}{\nu_1},\\
    b_{12} &=& \frac{\alpha_1\,(1-\epsilon_c)\,\overline{m}_0}{2\nu_1}, \\
    b_{21} &=& 0, \\
    b_{22} &=& \frac{\alpha_1\, \epsilon_c\,\overline{m}_0}{\nu_1}.
\end{eqnarray*}
The eigenvalues are then $b_{11}$ and $b_{22}$, which are equal to $1$ and $\epsilon_c$, respectively, since $\overline{m}_0 = \frac{\nu_1}{\alpha_1}$. 
Thus, we can conclude that  $R_I=1$, given $\epsilon_c \in [0,1]$.

\section{Alizon's proposal}
\label{app:alizon}

Alizon~\cite{alizon2013parasite} proposed a model
with doubly infected hosts 
(with the same viral strain),
and which seemed a  sufficient approach to achieve neutrality. 
We have considered a generalisation of the model originally
proposed in Ref.~\cite{alizon2013parasite}, and which is
described by the ODE system~\eqref{eq:alizon_model}. By setting $\frac{\dd m_0}{\dd t}=\frac{\dd m_1}{\dd t}=\frac{\dd M_1}{\dd t}=0$, and $m_2=m_c=M_2=0$, one obtains the endemic equilibrium for the resident strain $\widetilde{E}_1=(\widetilde{m}_0, \widetilde{m}_1, 0, 0,\widetilde{M}_1, 0)$. We then compute the invasion reproduction number of the invasive strain by considering the invasive sub-system, linearised around the resident strain endemic equilibrium, $\widetilde{E}_1$:
\begin{equation*}
\begin{split}
     \frac{\dd m_2}{\dd t} &= -\nu_2\,m_2 + \widetilde{m}_0\,\lambda_2 - m_2\,(\alpha_1\,\widetilde{m}_1 + 2\,\kappa_1\,\widetilde{M}_1),\\
     \frac{\dd m_c}{\dd t} &= -\nu_c\,m_c + \widetilde{m}_0\,(\lambda_{1,c} + \lambda_{2,c}) + \widetilde{m}_1\,(\lambda_2 + \lambda_{2,c} + \Lambda_2) + m_2\,(\alpha_1\,\widetilde{m}_1 + 2\,\kappa_1\,\widetilde{M}_1),\\
      \frac{\dd M_2}{\dd t} &= -\upsilon_2\,M_2 + \widetilde{m}_0\,\Lambda_2. 
\end{split}
\end{equation*}
The Jacobian matrix of the invasive sub-system is
\begin{eqnarray*}
    \widetilde{J} \equiv
\begin{pmatrix}
\substack{- \nu_2 + \widetilde{m}_0\,\alpha_2 \\ - (\alpha_1\,\widetilde{m}_1 + 2\,\kappa_1\,\widetilde{M}_1)}
& \substack{\widetilde{m}_0\,\delta_2\,(1- \epsilon_c)}
& \substack{2\,\widetilde{m}_0\,\kappa_2\,(1-\epsilon_2)}
\\ \addlinespace
\substack{\widetilde{m}_1\,\alpha_2 \\ + \alpha_1\,\widetilde{m}_1 + 2\,\kappa_1\,\widetilde{M}_1}
&
\substack{-\nu_c + \widetilde{m}_0\,(\delta_1+\delta_2)\,\epsilon_c \\ + \widetilde{m}_1\,\delta_2}
&
\substack{2\,\widetilde{m}_1\,\kappa_2}
\\ \addlinespace
\substack{0}
& \substack{0}
& \substack{-\upsilon_2+2\,\widetilde{m}_0\,\kappa_2\,\epsilon_2}
\end{pmatrix}
\; , 
\end{eqnarray*}
and can be decomposed as follows
\begin{eqnarray*}
    \widetilde{T} \equiv \begin{pmatrix}
        \widetilde{m}_0\,\alpha_2 
        & \widetilde{m}_0\,\delta_2\,(1-\epsilon_c) 
        & 2\,\widetilde{m}_0\,\kappa_2\,(1-\epsilon_2) 
        \\
        \widetilde{m}_1\,\alpha_2 
        & \widetilde{m}_0\,(\delta_1+\delta_2)\,\epsilon_c + \widetilde{m}_1\,\delta_2 
        & 2\,\widetilde{m}_1\,\kappa_2 
        \\
        0 
        & 0 
        & 2\,\widetilde{m}_0\,\kappa_2\,\epsilon_2
    \end{pmatrix}, \quad\text{and}
\end{eqnarray*}

\begin{eqnarray*}
    \widetilde{V} \equiv \begin{pmatrix}
       -\nu_2 - (\alpha_1\,\widetilde{m}_1 + 2\,\kappa_1\,\widetilde{M}_1)
        & 0
        & 0
        \\
        \alpha_1\,\widetilde{m}_1 + 2\,\kappa_1\,\widetilde{M}_1
        & -\nu_c
        & 0
        \\
        0 
        & 0 
        & -\upsilon_2
    \end{pmatrix}.
\end{eqnarray*}
Finally, the next-generation matrix, $\mathbb{K} \equiv \widetilde{T} [-\widetilde{V}]^{-1}$, is given by
\begin{eqnarray*}
\mathbb{K} \equiv \begin{pmatrix}
    A & B & C \\
    D & E & F \\
    G & H & I
\end{pmatrix}
\; ,
\end{eqnarray*}
where
\begin{eqnarray*}
    A &=& \frac{\alpha_2\,\widetilde{m}_0}{\nu_2 + \alpha_1\widetilde{m}_1 + 2\kappa_1 \widetilde{M}_1} + \frac{\alpha_1\widetilde{m}_1 + 2\kappa_1 \widetilde{M}_1}{\nu_2 + \alpha_1\widetilde{m}_1 + 2\kappa_1 \widetilde{M}_1}\,\frac{\delta_2\,(1-\epsilon_c)\,\widetilde{m}_0}{\nu_c}, \\
    B &=& \frac{\delta_2\,(1-\epsilon_c)\,\widetilde{m}_0}{\nu_c}, \\
    C &=& \frac{2\kappa_2(1-\epsilon_2)\,\widetilde{m}_0}{\upsilon_2}, \\
    D &=& \frac{\alpha_2\, \widetilde{m}_1}{\nu_2 + \alpha_1\widetilde{m}_1 + 2\kappa_1 \widetilde{M}_1} + \frac{\alpha_1\widetilde{m}_1 + 2\kappa_1 \widetilde{M}_1}{\nu_2 + \alpha_1\widetilde{m}_1 + 2\kappa_1 \widetilde{M}_1} \, \frac{(\delta_1+\delta_2)\, \epsilon_c\,\widetilde{m}_0 + \delta_2\,\widetilde{m}_1}{\nu_c}, \\
    E &=& \frac{(\delta_1+\delta_2)\, \epsilon_c\,\widetilde{m}_0 + \delta_2\,\widetilde{m}_1}{\nu_c},\\
    F &=& \frac{2\kappa_2\,\widetilde{m}_1}{\upsilon_2},\\
    G &=& 0,\\
    H &=& 0,\\
    I &=& \frac{2\kappa_2\epsilon_2\,\widetilde{m}_0}{\upsilon_2}.
\end{eqnarray*}
The invasion reproduction number, $R_I$, is given by the largest eigenvalue of $\mathbb{K}$. One eigenvalue is given by the matrix element $I$. The other eigenvalues are those of the sub-matrix
\begin{eqnarray*}
\mathbb{K}_{2\times 2} \equiv \begin{pmatrix}
    A & B\\
    D & E\\
\end{pmatrix}
\; ,
\end{eqnarray*}
with the largest of the two given by
\begin{equation*}
    R_I = \frac{(A+E) + \sqrt{(A+E)^2-4\,(A\,E-B\,D)}}{2}.
\end{equation*}

\subsection{Proof of neutrality}
\label{sec:neutrality_alizon}

When considering neutrality (\emph{i.e.,} the invasive strain is the same as the resident strain), we set $\nu_2=\nu_1=\nu$, $\upsilon_2=\upsilon_1=\upsilon$, $\alpha_2=\alpha_1=\alpha$, $\delta_2=\kappa_2=\delta_1=\kappa_1=\kappa$, and $\epsilon_c=\epsilon_2=\epsilon_1=\epsilon$. Here we will show that under these neutrality conditions, $R_I=1$, for the particular case where $\nu_0 = \upsilon = \nu$ and $\kappa = \alpha/2$. We make the simplifying assumption that infection does not affect the death rates since infection with CCHFV is asymptomatic in its animal hosts. The assumption that $\kappa = \alpha/2$ is realistic because we are considering viral infection. As Alizon mentioned in Ref.~\cite{alizon2013parasite}, if a host infected by a given strain is re-infected by the exact same strain, we do not expect to see a change in viral load (and hence in transmission rate); that is, a singly infected tick can become doubly infected with the same strain in this model, but becoming doubly infected does not affect its viral load. Under these conditions, the endemic equilibrium for the resident strain satisfies
\begin{equation}
\label{sum}
    \widetilde{m}_0 + \widetilde{m}_1 + \widetilde{M}_1 = \frac{\Phi}{\nu},
\end{equation}
with
\begin{eqnarray}
    \widetilde{m}_0 &=& \frac{\nu}{\alpha}, \label{m0}\\
    \widetilde{m}_1 &=& \frac{\widetilde{m}_0(1-\epsilon)(\alpha\Phi - \nu^2)}{\alpha\Phi - \nu^2\epsilon}. \label{m1}
\end{eqnarray}
Hence, we have
\begin{equation}\label{denom}
\nu + \alpha(\widetilde{m}_1 + \widetilde{M}_1) = \frac{\alpha\Phi}{\nu}.
\end{equation}
Making use of Eqs.~\eqref{sum}, \eqref{m0}, and \eqref{denom}, the relevant elements of the next-generation matrix simplify to
\begin{eqnarray*}
    A &=& \frac{\alpha\,\widetilde{m}_0}{\nu + \alpha(\widetilde{m}_1 + \widetilde{M}_1)} + \frac{\alpha(\widetilde{m}_1 + \widetilde{M}_1)}{\nu + \alpha(\widetilde{m}_1 + \widetilde{M}_1)}\,\frac{\alpha\,(1-\epsilon)\,\widetilde{m}_0}{2\nu}, \\
    &=& \frac{1}{\Phi}\left[
    \nu\,\widetilde{m}_0 + \frac{1}{2}\nu(1-\epsilon)\left(\frac{\Phi}{\nu} -\widetilde{m}_0\right)\right],
    \\
    B &=& \frac{\alpha\,(1-\epsilon)\,\widetilde{m}_0}{2\nu}
    = \frac{1}{2}(1-\epsilon),
    \\
    D &=& \frac{\alpha\, \widetilde{m}_1}{\nu + \alpha(\widetilde{m}_1 + \widetilde{M}_1)} + \frac{\alpha(\widetilde{m}_1 + \widetilde{M}_1)}{\nu + \alpha(\widetilde{m}_1 + \widetilde{M}_1)} \, \frac{\alpha\, \epsilon\,\widetilde{m}_0 + \frac{1}{2}\alpha\,\widetilde{m}_1}{\nu}, \\
    &=& \frac{1}{\Phi}\left[
    \nu\, \widetilde{m}_1 + \nu\left(\epsilon + \frac{\alpha}{2\nu}\,\widetilde{m}_1\right)\left(\frac{\Phi}{\nu} -\widetilde{m}_0\right)\right], 
    \\
    E &=& \frac{\alpha\, \epsilon\,\widetilde{m}_0 + \frac{1}{2}\alpha\,\widetilde{m}_1}{\nu} =
    \epsilon + \frac{\alpha}{2\nu}\,\widetilde{m}_1,\\
    I &=& \frac{\alpha\epsilon\,\widetilde{m}_0}{\nu}=
    \epsilon.
\end{eqnarray*}
Since $I=\epsilon\leq 1$, we need to show that
\begin{equation*}
    R_I = \frac{(A+E) + \sqrt{(A+E)^2-4\,(A\,E-B\,D)}}{2} = 1.
\end{equation*}
It can be shown that this is true if and only if
\begin{equation}\label{neutrality_cond}
    A + E - (AE-BD) = 1.
\end{equation}
Thus, in order to show that $R_I=1$, it is sufficient to show that 
Eq.~\eqref{neutrality_cond} holds. We have,
$$AE-BD = \frac{1}{\Phi}\epsilon\nu\left(\widetilde{m}_0 + \frac{1}{2}\widetilde{m}_1\right).$$
Thus, we can write
\begin{equation*}
    \begin{split}
        A + E - (AE-BD) &= \frac{1}{\Phi}\left[
        \nu\,\widetilde{m}_0 + \frac{1}{2}\nu(1-\epsilon)\left(\frac{\Phi}{\nu} -\widetilde{m}_0\right) + \Phi\epsilon + \frac{\alpha\Phi}{2\nu}\,\widetilde{m}_1 - \epsilon\nu\left(\widetilde{m}_0 + \frac{1}{2}\widetilde{m}_1\right)\right],
        \\
        &= \frac{1}{\Phi}\left[
        \frac{1}{2}\nu\,\widetilde{m}_0(1-\epsilon) + \frac{\Phi}{2}\left(1+\epsilon\right) + \frac{1}{2\nu}\widetilde{m}_1(\alpha\Phi- \nu^2\epsilon)\right].\\
    \end{split}
\end{equation*}
Substituting $\widetilde{m}_1$ with its expression from Eq.~\eqref{m1} gives
\begin{equation*}
    \begin{split}
        A + E - (AE-BD) &= \frac{1}{\Phi}\left[
        \frac{\alpha\Phi}{2\nu}\,\widetilde{m}_0(1-\epsilon) + \frac{\Phi}{2}\left(1+\epsilon\right)\right].\\
    \end{split}
\end{equation*}
Finally, by substituting Eq.~\eqref{m0} in the previous equation, we have 
$$A + E - (AE-BD) = 1.$$

\section{The two-slot model of co-infection and co-transmission}
\label{app:two-slot}

\subsection{Transmission events}

We list here the transmission events which lead to the two-slot mathematical model introduced in Eq.~\eqref{eq:two-slot} grouped by the type of transmission.
$T_0$ denotes a susceptible tick,
$T_1$ and $T_2$ denote a singly infected tick with $V_1$
and $V_2$, respectively,
 $T_{11}$ and $ T_{22}$
 denote a doubly infected tick with $V_1$
and $V_2$, respectively,
and $T_c$ denotes a co-infected tick (with  $V_1$
and $V_2$).

\begin{itemize}
    \item Transmission from a singly infected tick to a susceptible tick:

$T_0 + T_1 \to T_1 + T_1$ with rate $\alpha_1 m_0 m_1$,

$T_0 + T_2 \to T_2 + T_2$ with rate $\alpha_2 m_0 m_2$.

\item Transmission from a singly infected tick to a singly infected tick:

$T_1 + T_1 \to T_{11} + T_1$ with rate $\alpha_1 m_1 m_1$,

$T_1 + T_2 \to T_c + T_2$ with rate $\alpha_2 m_1 m_2$,

$T_2 + T_1 \to T_c + T_1$ with rate $\alpha_1 m_2 m_1$,

$T_2 + T_2 \to T_{22} + T_2$ with rate $\alpha_2 m_2 m_2$. 

\item Transmission from a doubly infected tick to a susceptible tick:

$T_0 + T_{11} \to T_1 + T_{11}$ with rate $(1-\epsilon_1) 2 \kappa_1 m_0 M_1$,

$T_0 + T_{11} \to T_{11} + T_{11}$ with rate $\epsilon_1 2 \kappa_1 m_0 M_1$. 

$T_0 + T_{22} \to T_2 + T_{22}$ with rate $(1-\epsilon_2) 2 \kappa_2 m_0 M_2$,

$T_0 + T_{22} \to T_{22} + T_{22}$ with rate $\epsilon_2 2 \kappa_2 m_0 M_2$. 

\item Transmission from a co-infected tick to a susceptible tick:

$T_0 + T_c \to T_1 +T_c$ with rate $(1-\epsilon_c) \delta_1 m_0 m_c$,

$T_0 + T_c \to T_2 +T_c$ with rate $(1-\epsilon_c) \delta_2 m_0 m_c$,

$T_0 + T_c \to T_c +T_c$ with rate $\frac{2 \delta_1 \delta_2}{\delta_1 + \delta_2} \epsilon_c m_0 m_c$,

$T_0 + T_c \to T_{11} +T_c$ with rate $\frac{\delta_1^2}{\delta_1 + \delta_2} \epsilon_c m_0 m_c$,

$T_0 + T_c \to T_{22} +T_c$ with rate $\frac{\delta_2^2}{\delta_1 + \delta_2} \epsilon_c m_0 m_c$. 

\item Transmission from a co-infected tick to a singly infected tick:

$T_1 + T_c \to T_{11} + T_c$ with rate $\delta_1 m_1 m_c$, 

$T_1 + T_c \to T_c + T_c$ with rate $\delta_2 m_1 m_c$.

$T_2 + T_c \to T_{22} + T_c$ with rate $\delta_2 m_2 m_c$, 

$T_2 + T_c \to T_c + T_c$ with rate $\delta_1 m_2 m_c$. 

\item Transmission from a doubly infected tick to a singly infected tick:

$T_1 + T_{11} \to T_{11} + T_{11}$ with rate $2 \kappa_1 m_1 M_1$,

$T_2 + T_{22} \to T_{22} + T_{22}$ with rate $2 \kappa_2 m_2 M_2$,

$T_1 + T_{22} \to T_c + T_{22}$ with rate $2 \kappa_2 m_1 M_2$,

$T_2 + T_{11} \to T_c + T_{11}$ with rate $2 \kappa_1 m_2 M_1$.

\end{itemize}

\subsection{Existence of the endemic equilibrium of $V_1$}

In \ref{sec:neutrality_alizon} we have shown the endemic equilibrium (EE) of $V_1$ can be written as $\widetilde{E}_1=(\widetilde{m}_0, \widetilde{m}_1, 0,0,\widetilde{M}_1,0)$.
We have also shown  the neutrality of the invasion reproduction number in Alizon's model under the assumption $\kappa=\alpha/2$. This assumption simplifies the next-generation matrix and helps to prove  neutrality. Our two-slot model, as an extension of Alizon's model, shares the same resident strain EE. In this section, we will discuss the existence of the resident strain EE when $\kappa \neq \alpha/2$, referring to~\ref{sec:neutrality_alizon} for the case $\kappa = \alpha/2$.  We, thus, write the EE as $E_1^\prime=(m_0^\prime, m_1^\prime, 0, 0, M_1^\prime, 0)$. We then compute the invasion reproduction number $R_I^\prime$ and prove the neutrality in \ref{sect:two_slot_RI} and \ref{sect:two_slot_neutrality}, respectively.

First, we consider the resident sub-system, where $m_2=m_C=M_2=0$ and assume $\nu_0=\nu_1=\nu_2=\nu_c=\upsilon_1=\upsilon_2=\nu$, without loss of generality.
We have
\begin{equation}
    \begin{split}
        \frac{\dd m_0}{\dt} & = \Phi - \nu\, m_0 - 
       (\alpha_1\,m_1 + 2\,\kappa_1\,M_1)\,m_0, \\
        \frac{\dd m_1}{\dt} & = - \nu\,m_1 - 
        (\alpha_1\,m_1 + 2\,\kappa_1\,M_1)\,m_1 + (\alpha_1\,m_1+2\,\kappa_1\,(1-\epsilon_1)\,M_1)\,m_0, \\
        \frac{\dd M_1}{\dt} & = - \nu\,M_1 + (\alpha_1\,m_1 + 2\,\kappa_1\,M_1)\,m_1 + 2\,\epsilon_1\,\kappa_1\,M_1\,m_0.
    \end{split}
\end{equation}
We compute the basic reproduction number of this sub-system at the VFE $E_0=(m_0^\star,0,0,0,0,0)$, where $m_0^\star=\frac{\Phi}{\nu}$. We can write
\begin{align}
    R_1^\prime &= \max\left\{\frac{\Phi\,\alpha_1}{\nu^2},\,\frac{2\,\Phi\,\epsilon_1\,\kappa_1}{\nu^2}\right\}.
\end{align}
By setting $\frac{\dd (m_1+M_1)}{\dd t}=0$ and $\frac{\dd m_0}{\dd t}=0$, we obtain the following equations:
\begin{align}
     \alpha_1\,m_1^\prime + 2\,\kappa_1\,M_1^\prime&= \frac{(m_1^\prime + M_1^\prime)\,\nu}{m_0^\prime}, \label{eqn:e1}\\
     m_1^\prime + M_1^\prime &= \frac{\Phi}{\nu} - m_0^\prime. \label{eqn:e2}
\end{align}
From Eq.~\eqref{eqn:e2}, we have $m_0^\prime < \frac{\Phi}{\nu}$ to ensure positive $m_1^\prime$ and $M_1^\prime$. By combining Eqs.~\eqref{eqn:e1} and \eqref{eqn:e2}, we then get
\begin{align}
    \alpha_1\,m_1^\prime + 2\,\kappa_1\,M_1^\prime &= \frac{\Phi}{m_0^\prime} - \nu.
    \label{eqn:e3}
\end{align}
By solving Eq.~\eqref{eqn:e2} and Eq.~\eqref{eqn:e3}, one obtains
\begin{equation}
\label{eqn:expr_m1M1}
    m_1^\prime = \frac{\Phi-\nu\,m_0^\prime}{2\kappa_1-\alpha_1}\left(\frac{1}{m_0^\prime}-\frac{\alpha_1}{\nu}\right)
    \; ,
    \quad \text{and} \;  \quad
    M_1^\prime = \frac{\Phi-\nu\,m_0^\prime}{\alpha_1 - 2\kappa_1}\left(\frac{1}{m_0^\prime}-\frac{2\kappa_1}{\nu}\right)
    \; .
\end{equation}
We conclude that to ensure
 positive values for $m_1^\prime$ and $M_1^\prime$, we require the following conditions: 
\begin{itemize}
    \item if $\alpha_1 >2\kappa_1$, $\frac{\nu}{\alpha_1}<m_0^\prime<\frac{\nu}{2\kappa_1}$, or 
    \item if $\alpha_1<2\kappa_1$, $\frac{\nu}{2\kappa_1} < m_0^\prime < \frac{\nu}{\alpha_1}$.
\end{itemize}
If $2\kappa_1 = \alpha_1$, we refer to \ref{sec:neutrality_alizon}.
Substituting Eqs.~\eqref{eqn:e3} and \eqref{eqn:expr_m1M1} into $\frac{\dd M_1}{\dd t}=0$, we derive the following cubic equation for $m_0^\prime$:
\begin{equation*}
    Q(m_0)=A_3\,m_0^3 + A_2\,m_0^2 + A_1\,m_0 + A_0=0,
\end{equation*}
where $A_3=-2\kappa_1\,\epsilon_1\,\alpha_1,\,A_2=2\kappa_1\,\epsilon_1\,\nu - 2\kappa_1\,\nu + \alpha_1\,\nu,\,A_1=2\Phi\,\kappa_1$, and $A_0=-\Phi\,\nu$.
It is easy to observe that $A_3 < 0,\,Q(0)<0$, and $Q^\prime(0)>0$.
We therefore have one negative root for $Q(m_0^\prime)=0$ and two critical points ({\em i.e.,} where $Q^\prime(m_0^\prime) = 0$) distributed at different sides of the $y$-axis. 
From Eqs.~\eqref{eqn:e2} and \eqref{eqn:expr_m1M1}, we have three important values for $m_0^\prime$: $\frac{\Phi}{\nu}$, $\frac{\nu}{\alpha_1}$, and $\frac{\nu}{2\kappa_1}$. We then have the following values:
\begin{align*}
    Q\left(\frac{\Phi}{\nu}\right) &= \Phi\,\nu\,\left(1-\frac{\Phi}{\nu}\frac{2\,\kappa_1\,\epsilon_1}{\nu}\right)\left(\frac{\Phi}{\nu}\frac{\alpha_1}{\nu}-1\right), \\
    Q\left(\frac{\nu}{\alpha_1}\right) &= \nu^2\,\left(\frac{\Phi}{\nu}-\frac{\nu}{\alpha_1}\right)\left(\frac{2\,\kappa_1}{\alpha_1}-1\right), \\
    Q\left(\frac{\nu}{2\,\kappa_1}\right) &= \frac{\nu^3}{2\,\kappa_1}\,\left(\frac{\alpha_1}{2\,\kappa_1}-1\right)\left(1-\epsilon_1\right).
\end{align*}
We can now discuss the existence of a real and positive $m_0^\prime$, when $R_1^\prime >1$ and $0\leq \epsilon_1\leq 1$.
We need to consider the following cases:
\begin{enumerate}[(i)]
    \item 
    when $\frac{2\,\kappa_1\,\epsilon_1\,\Phi}{\nu^2}\leq 1 <\frac{\alpha_1\Phi}{\nu^2}$, then $R_1^\prime=\frac{\alpha_1\Phi}{\nu^2}$, $Q\left(\frac{\Phi}{\nu}\right)\geq 0$ and $\frac{\nu}{\alpha_1} < \frac{\Phi}{\nu}$, $\frac{\nu}{2\kappa_1} \geq \epsilon_1\frac{\Phi}{\nu}$.
    We need to consider two separate cases:
    
    \begin{enumerate}[(a)]
        \item if $\alpha_1 > 2\kappa_1$, 
        that is, $\frac{\nu}{\alpha_1}<m_0^\prime<\frac{\nu}{2\kappa_1}$, then $Q\left(\frac{\nu}{\alpha_1}\right)<0$, and $Q\left(\frac{\nu}{2\,\kappa_1}\right)\geq 0$. We thus have a unique solution for $Q(m_0^\prime)=0$ on $\left(\frac{\nu}{\alpha_1},\min\left\{\frac{\nu}{2\kappa_1},\frac{\Phi}{\nu}\right\}\right]$.
        
        \item if $\alpha_1 < 2\kappa_1$, that is, $\frac{\nu}{2\kappa_1} < m_0^\prime < \frac{\nu}{\alpha_1}$ and $2\kappa_1\epsilon_1<\alpha_1<2\kappa_1$, then $Q\left(\frac{\nu}{\alpha_1}\right)>0$ and $Q\left(\frac{\nu}{2\,\kappa_1}\right)\leq 0$. We thus have a unique solution on $\left[\frac{\nu}{2\kappa_1},\frac{\nu}{\alpha_1}\right)$.
        
    \end{enumerate}
    
    \item when $\frac{\alpha_1\Phi}{\nu^2} \leq 1 <\frac{2\,\kappa_1\,\epsilon_1\,\Phi}{\nu^2}$, then $R_1^\prime=\frac{2\,\kappa_1\,\epsilon_1\,\Phi}{\nu^2}$, $\alpha_1 < 2\kappa_1\epsilon_1 < 2\kappa_1$, $\frac{\nu}{2\kappa_1}<\epsilon_1\frac{\Phi}{\nu}$, and $\frac{\nu}{\alpha_1}\geq \frac{\Phi}{\nu}$,
    since $\alpha_1 < 2\kappa_1$; that is, $\frac{\nu}{2\kappa_1}<m_0^\prime<\frac{\nu}{\alpha_1}$. Then we can further constrain the solution to $\frac{\nu}{2\kappa_1}<m_0^\prime<\frac{\Phi}{\nu}$, and we have $Q\left(\frac{\nu}{2\kappa_1}\right)\leq 0$,
    and $Q\left(\frac{\Phi}{\nu}\right)> 0$. Thus a unique solution can be found on $\left[\frac{\nu}{2\kappa_1},\frac{\Phi}{\nu}\right)$.
    
    \item when $\frac{\alpha_1\Phi}{\nu^2}>1$ and $\frac{2\kappa_1\epsilon_1\Phi}{\nu^2}>1$, then $\frac{\nu}{\alpha_1}<\frac{\Phi}{\nu}$, $\frac{\nu}{2\kappa_1}<\epsilon_1\frac{\Phi}{\nu}$, and $Q\left(\frac{\Phi}{\nu}\right)< 0$.
    We need to consider two different cases:
    
    \begin{enumerate}[(a)]
        \item if $\alpha_1 > 2\kappa_1$, that is $\frac{\nu}{\alpha_1}<m_0^\prime<\frac{\nu}{2\kappa_1}$, then $Q\left(\frac{\nu}{\alpha_1}\right) < 0$ and $Q\left(\frac{\nu}{2\kappa_1}\right)\geq 0$. 
        We thus have a unique solution on $\left(\frac{\nu}{\alpha_1},\frac{\nu}{2\kappa_1}\right]$.
        
        \item if $\alpha_1 < 2\kappa_1$, that is $\frac{\nu}{2\kappa_1} < m_0^\prime < \frac{\nu}{\alpha_1}$, then $Q\left(\frac{\nu}{\alpha_1}\right)>0$ and $Q\left(\frac{\nu}{2\kappa_1}\right)\leq 0$. We thus have a unique solution on $\left[\frac{\nu}{2\kappa_1},\frac{\nu}{\alpha_1}\right)$.
    \end{enumerate}
\end{enumerate}
Therefore, when $R_1^\prime>1$ and $0\leq\epsilon_1\leq 1$, a unique real and non-negative solution of $E_1^\prime=(m_0^\prime, m_1^\prime, 0, 0, M_1^\prime, 0)$ is guaranteed.
We can ensure three real roots ({\em i.e.,} one negative and two positive roots) for $Q(m_0)=0$, such that we identify the value of $m_0^\prime$ which is real and  positive, with $m_1^\prime$ and $M_1^\prime$  real and  positive as well.
To do so we make use of the general formula for a cubic equation (with $A_3\neq0$)~\cite{cardano2007rules}:
\begin{align}
\label{eqn:m0}
    m_0^\prime &= -\frac{1}{3\,A_3}\,\left(A_2 + \Delta + \frac{\Delta_0}{\Delta}\right),
\end{align}
with
\begin{align*}
    \Delta_0 &= A_2^2 - 3\,A_1\,A_3, 
    \\
    \Delta_1 &= 2\,A_2^3 - 9\,A_1\,A_2\,A_3 + 27\,A_0\,A_3^2, 
    \\
    \Delta &= \left(\frac{-1+\sqrt{-3}}{2}\right)^2\,\left(\frac{\Delta_1 + \sqrt{\Delta_1^2 - 4\,\Delta_0^3}}{2}\right)^{1/3}.
\end{align*}
We note that 
 $m_0^\prime$ is real even though  $\Delta$ is a complex number. Another expression 
  for the solution of 
  $m_0^\prime$  making use of trigonometric functions 
 can be found in Ref.~\cite{zwillinger2018crc}.

\subsection{Invasion reproduction number}
\label{sect:two_slot_RI}

We can now compute the invasion reproduction number, $R_I^\prime$, of 
 $V_2$ for the invasion sub-system, linearised around the endemic equilibrium $E_1^\prime$.
We have the following Jacobian matrix:
\begin{equation*}
    J^\prime \equiv \begin{pmatrix}
        \substack{-\nu-\alpha_1\,m_1^\prime \\+ \alpha_2\,m_0^\prime - 2\,\kappa_1\,M_1^\prime} & \delta_2\,(1-\epsilon_c)\,m_0^\prime & 2\,\kappa_2\,(1-\epsilon_2)\,m_0^\prime \\ \addlinespace
        \substack{(\alpha_1 + \alpha_2)\,m_1^\prime \\+ 2\,\kappa_1\,M_1^\prime} & -\nu+\delta_2\,\left(\frac{2\delta_1\,\epsilon_c}{\delta_1+\delta_2}\,m_0^\prime + m_1^\prime\right) & 2\,\kappa_2\,m_1^\prime \\ \addlinespace
        0 & \frac{\delta_2^2\,\epsilon_c}{\delta_1+\delta_2}\,m_0^\prime & -\nu+2\,\epsilon_2\,\kappa_2\,m_0^\prime
    \end{pmatrix},
\end{equation*}
which can be decomposed as follows
\begin{equation*}
    T^\prime \equiv \begin{pmatrix}
        \alpha_2\,m_0^\prime & \delta_2\,(1-\epsilon_c)\,m_0^\prime & 2\,\kappa_2\,(1-\epsilon_2)\,m_0^\prime \\
        \alpha_2\,m_1^\prime & \delta_2\left(\frac{2\delta_1\,\epsilon_c}{\delta_1+\delta_2}\,m_0^\prime + m_1^\prime\right) & 2\,\kappa_2\,m_1^\prime \\
        0 & \frac{\delta_2^2\,\epsilon_c}{\delta_1+\delta_2}\,m_0^\prime & 2\,\epsilon_2\,\kappa_2\,m_0^\prime
    \end{pmatrix},
\end{equation*}
and 
\begin{equation*}
    V^\prime \equiv \begin{pmatrix}
        -\nu-\alpha_1\,m_1^\prime - 2\,\kappa_1\,M_1^\prime & 0 & 0 \\
        \alpha_1\,m_1^\prime + 2\,\kappa_1\,M_1^\prime & -\nu & 0 \\
        0 & 0 & -\nu
    \end{pmatrix}.
\end{equation*}
We can compute the next-generation matrix, $\mathbb{K}^\prime=T^\prime[-V^\prime]^{-1}$, given by:
\begin{equation*}
    \mathbb{K}^\prime \equiv \begin{pmatrix}
        d_{11} & d_{12} & d_{13} \\
        d_{21} & d_{22} & d_{23} \\
        d_{31} & d_{32} & d_{33} 
    \end{pmatrix},
\end{equation*}
where
\begin{align*}
    d_{11} &= \frac{\alpha_2\,m_0^\prime+\delta_2\,(1-\epsilon_c)\,(m_1^\prime + M_1^\prime)}{\Phi}\,m_0^\prime, \\
    d_{12} &= \frac{\delta_2}{\nu}\,(1-\epsilon_c)\,m_0^\prime, \\
    d_{13} &= \frac{2\,\kappa_2}{\nu}\,(1-\epsilon_2)\,m_0^\prime, \\
    d_{21} &= \frac{\alpha_2}{\Phi}\,m_0^\prime\,m_1^\prime + \frac{\delta_2}{\Phi}\left(\frac{2\,\delta_1\,\epsilon_c}{\delta_1+\delta_2}m_0^\prime + m_1^\prime\right)\left(m_1^\prime + M_1^\prime\right), \\
    d_{22} &= \frac{\delta_2}{\nu}\,\left(\frac{2\,\epsilon_c\,\delta_1}{\delta_1+\delta_2}\,m_0^\prime+m_1^\prime\right), \\
    d_{23} &= \frac{2\,\kappa_2}{\nu}\,m_1^\prime, \\
    d_{31} &= \frac{\delta_2}{\Phi}\,\frac{\epsilon_c\,\delta_2}{\delta_1+\delta_2}\,m_0^\prime\left(m_1^\prime+M_1^\prime\right), \\
    d_{32} &= \frac{\delta_2}{\nu}\,\frac{\epsilon_c\,\delta_2}{\delta_1+\delta_2}\,m_0^\prime, \\
    d_{33} &= \frac{2\,\epsilon_2\,\kappa_2}{\nu}\,m_0^\prime. \\
\end{align*}
We can obtain the invasion reproduction number, $R_I^\prime$, as a function of $m_0^\prime$ by substituting Eq.~\eqref{eqn:expr_m1M1} into $\mathbb{K}^\prime$ and computing its largest eigenvalue.

\subsection{Proof of neutrality}
\label{sect:two_slot_neutrality}
We consider the following limits: $\alpha_1, \alpha_2\to\alpha$, $\kappa_1,\kappa_2,\delta_1,\delta_2\to\kappa$, and $\epsilon_1, \epsilon_2, \epsilon_c\to\epsilon$.
In this limit, we can write  the
 invasion reproduction number, $R_I^\prime$, at neutrality as follows:
\begin{align}
\begin{split}
    R_I^\prime = & \,\, \frac{m_0^\prime}{2\,\nu\,\Phi\,\left(\Phi-\alpha\,\epsilon\,{m_0^\prime}^2\right)}\biggl\{\\
    &\quad 2\,\Phi^2\,\kappa+\Phi\,\nu\,(\alpha-2\,\kappa\,(1-\epsilon))\,m_0^\prime \\
    &\quad -\Phi\,\alpha\,\epsilon\,\kappa\,(1+\epsilon){m_0^\prime}^2-\alpha\,\epsilon\,\nu\,(\alpha-\kappa\,(1-\epsilon))\,{m_0^\prime}^3 \\
    &\quad +\Bigr[4\,\Phi\,\alpha\,\epsilon\,\kappa\,\nu\,m_0^\prime\,\left(\Phi-\alpha\,\epsilon\,{m_0^\prime}^2\right) \Bigl(\Phi\,(-3+\epsilon) +m_0^\prime\,\left(\nu\,(1-\epsilon) +2\,\alpha\,\epsilon\,m_0^\prime\right)\Bigl) \\
    &\quad\quad +\Bigl(2\,\Phi^2\,\kappa+m_0^\prime\bigr(\Phi\,\nu(\alpha-2\,\kappa(1-\epsilon))\\
    &\quad\quad\quad - \epsilon\,\alpha\,m_0^\prime((1+\epsilon)\,\Phi\,\kappa+\nu\,m_0^\prime\,(\alpha-(1-\epsilon)\,\kappa))\bigr)\Bigl)^2\Bigr]^{1/2}\biggl\}.
\end{split}
\end{align}
Now we can prove the neutrality of $R_I^\prime$ in two scenarios: 
\begin{enumerate}[(1)]
    \item when $\kappa=\alpha/2$, the expression for $R_I^\prime$  reduces to
        \begin{align*}
            R_I^\prime =& \frac{\alpha\,m_0^\prime\left(2\Phi\,(\Phi + \epsilon\,\nu\,m_0^\prime)-\epsilon\,\alpha(1+\epsilon)(\Phi+\nu\,m_0^\prime){m_0^\prime}^2\right)}{4\,\Phi\,\nu\,(\Phi-\epsilon\,\alpha\,{m_0^\prime}^2)} \\
            &+\,\,\alpha\,m_0^\prime\left[
                \frac{\epsilon\,m_0^\prime\,((2\epsilon\alpha\,m_0^\prime+(1-\epsilon)\nu)m_0^\prime-(3-\epsilon)\Phi)}{2\nu\,\Phi\,(\Phi-\epsilon\,\alpha\,{m_0^\prime}^2)}\right. \\
            &+\,\,\left.\left(\frac{2\Phi^2+\epsilon\,m_0^\prime\,(2\Phi\,\nu-\alpha\,m_0^\prime\,(1+\epsilon)\,(\Phi+\nu\,m_0^\prime))}{4\Phi\nu\,(\Phi-\epsilon\,\alpha\,{m_0^\prime}^2)}\right)^2\right]^{1/2}.
        \end{align*}
        By substituting $m_0^\prime=\widetilde{m}_0=\nu/\alpha$ as in Eq.~\eqref{m0}, we  obtain $R_I^\prime \equiv 1$, as desired. 
        
    \item when $\kappa\neq\alpha/2$, the expression  for 
    $R_I^\prime$ cannot be
    easily simplified. In this case,  we perform a numerical
    study, making use of Mathematica to prove  neutrality,
    which requires
  the expression of $m_0^\prime$ from Eq.~\eqref{eqn:m0}.
    
\end{enumerate}

\section*{Acknowledegements}

We thank
Dr. Jonathan Carruthers  (UKHSA) for preparing
 Figure~\ref{fig:reassortment},
 and Dr. Macauley Locke (LANL) for research
 discussions on model development and parameterisation.

\section*{Funding}

This work was supported by the 
Biotechnology and Biological Sciences Research Council
Research Council [grant number BB/W010755/1]
(B.W., Z.V., M.L.-G., and G.L.).
This study was supported
 by the National Institutes of Health/National Institute of Allergy and Infectious Diseases grant R01AI087520 to T.L.,
and grant 
R01AI167048 to E.R.-S., T.L. and C.M.-P.
This project has received funding from the European Union’s Horizon 2020 research and innovation program under the Marie Skłodowska-Curie Grant, agreement number 764698 (G.B., G.L. and C.M.-P.).
Y.L.'s research was partially supported by the NSF of China [12071393].

\section*{Data availability statement}

Numerical codes (Python) to reproduce
Figure~\ref{fig:SingleR0} and Figure~\ref{fig:heatmaps_RI}, as  well as the Mathematica notebook to reproduce 
proofs and results from~\ref{app:two-slot}, are deposited at \url{https://github.com/MolEvolEpid/coinfection_cotransmission_cofeeding_in_ticks}.



\section*{Selected references}

\begin{enumerate}

\item 
Maliyoni {\em et al.}~\cite{maliyoni2019stochastic} proposed a stochastic model for the dynamics of two tick-transmitted pathogens in a single tick population. The model, a continuous-time Markov chain based on a deterministic tick-borne disease model, was used to investigate the duration of possible pathogen co-existence and the probability of pathogen extinction.

\item
Cutler {\em et al.}~\cite{cutler2021tick} reviewed current understanding of co-infection in tick-borne diseases affecting both tick and vertebrate host populations, highlighting the need for more research on pathogen interactions.

\item 
Vogels {\em et al.}~\cite{vogels2019arbovirus} reviewed the
impact of co-infection on clinical disease in humans, discussed the possibility for co-transmission from mosquito to human, and described a role for modeling transmission dynamics at various levels of co-transmission with the aim of understanding whether virus co-infections should be viewed as a serious concern for public health.

\item 
Meehan {\em et al.}~\cite{meehan2020probability} 
investigated the stochastic dynamics of the
emergence of a novel disease strain, which is introduced into a population in which it competes with a resident endemic strain. The analysis is carried out by means of a branching
process approximation to calculate the probability that the new strain becomes established.

\item 
Allen {\em et al.}~\cite{allen2019modelling} 
formulated a general epidemiological model for one vector species and one plant species with potential co-infection in the host plant by two viruses. First, the basic reproduction number is derived, and thus, conditions for successful invasion of a single virus are determined. Then, a new invasion threshold is derived to provide conditions for successful invasion of a second virus.

\item 
White {\em et al.}~\cite{white2019dynamics} proposed a mathematical model for a two-pathogen, one-tick, one-host system with the aim of determining how long an invading pathogen persists within a tick population in which a resident pathogen is already established.

\item 
Rovenolt {\em et al.}~\cite{rovenolt2022impact} developed a model of two co-infected host species to understand under which conditions co-infection can interfere with parasite-mediated apparent competition among hosts.  

\item 
Le {\em et al.}~\cite{le2022disentangling} studied a Susceptible-Infected-Susceptible (SIS) compartmental model with two strains and co-infection/co-colonization, incorporating five strain fitness dimensions under the same framework to understand coexistence and competition mechanisms.

\item 
Alizon~\cite{alizon2013co} discussed how multiple infections have been modelled in evolutionary epidemiology, presenting within-host models, super-infection frameworks, co-infection models, and some perspectives for the study of multiple infections in evolutionary epidemiology. In particular, he showed that a widely used co-infection model is {\em not neutral} as it confers a frequency-dependent advantage to rare neutral mutants.

\item 
Alizon~\cite{alizon2013parasite} studied the effect of co-transmission on virulence evolution when parasites compete for host resources.

\item 
Bhowmick {\em et al.}~\cite{bhowmick2022ticks} developed a compartment-based non-linear ordinary differential equation system to model the disease transmission cycle including blood-sucking ticks, livestock
and humans. Sensitivity analysis of the basic reproduction number shows that decreasing the tick
survival time is an efficient method to control the disease.
They concluded that in the case of CCHFV 
transmission due to co-feeding, as well as 
trans-stadial and trans-ovarial transmission, are
important routes to sustain the disease cycle.

\item 
Hoch {\em et al.}~\cite{hoch2018dynamic} proposed 
a dynamic mechanistic model that takes into account the major processes involved in tick population dynamics and pathogen transmission with the aim of testing potential scenarios for pathogen control.

\item 
Johnstone {\em et al.}~\cite{johnstone2020incorporating} derived expressions for the basic reproduction number and the related tick type-reproduction number accounting for the observation that larval and nymphal ticks tend to aggregate on the same minority of hosts (tick co-aggregation and co-feeding). The pattern of tick blood meals is represented as a directed, acyclic, bipartite contact network.

\item 
Belluccini~\cite{belluccini2023stochastic} proposed both deterministic and stochastic models of co-infection with tick-borne viruses to investigate the role that different routes of transmission play in the spread of infectious diseases and to study the probability and timescale of co-infection events.

\item 
Maliyoni {\em et al.}~\cite{maliyoni2023multipatch} investigated the impact of between-patch migration on the dynamics of a tick-borne disease on disease extinction and persistence making use of a system of stochastic differential equations.

\item 
Lin {\em et al.}~\cite{lin2023co} studied the impact of Zika and Dengue virus co-infection on viral infection, examining viral replication activity in cells infected simultaneously, or sequentially.

\end{enumerate}

\bibliographystyle{elsarticle-num} 
\bibliography{cas-refs}

\begin{thebibliography}{10}
\expandafter\ifx\csname url\endcsname\relax
  \def\url#1{\texttt{#1}}\fi
\expandafter\ifx\csname urlprefix\endcsname\relax\def\urlprefix{URL }\fi
\expandafter\ifx\csname href\endcsname\relax
  \def\href#1#2{#2} \def\path#1{#1}\fi

\bibitem{mcdonald2016reassortment}
S.~M. McDonald, M.~I. Nelson, P.~E. Turner, J.~T. Patton, Reassortment in
  segmented {RNA} viruses: mechanisms and outcomes, Nature Reviews Microbiology
  14~(7) (2016) 448--460.

\bibitem{perez2015recombination}
M.~P{\'e}rez-Losada, M.~Arenas, J.~C. Gal{\'a}n, F.~Palero,
  F.~Gonz{\'a}lez-Candelas, Recombination in viruses: mechanisms, methods of
  study, and evolutionary consequences, Infection, Genetics and Evolution 30
  (2015) 296--307.

\bibitem{negredo2021fatal}
A.~Negredo, R.~S{\'a}nchez-Arroyo, F.~D{\'\i}ez-Fuertes, F.~De~Ory, M.~A.
  Budi{\~n}o, A.~V{\'a}zquez, {\'A}.~Garcinu{\~n}o, L.~Hern{\'a}ndez, C.~de~la
  Hoz~Gonz{\'a}lez, A.~Guti{\'e}rrez-Arroyo, et~al., Fatal case of
  {Crimean-Congo} hemorrhagic fever caused by reassortant virus, {Spain}, 2018,
  Emerging Infectious Diseases 27~(4) (2021) 1211.

\bibitem{gerrard2004ngari}
S.~R. Gerrard, L.~Li, A.~D. Barrett, S.~T. Nichol, Ngari virus is a
  {Bunyamwera} virus reassortant that can be associated with large outbreaks of
  hemorrhagic fever in {Africa}, Journal of virology 78~(16) (2004) 8922--8926.

\bibitem{cline2011increased}
T.~D. Cline, E.~A. Karlsson, P.~Freiden, B.~J. Seufzer, J.~E. Rehg, R.~J.
  Webby, S.~Schultz-Cherry, Increased pathogenicity of a reassortant 2009
  pandemic {H1N1} influenza virus containing an {H5N1} hemagglutinin, Journal
  of virology 85~(23) (2011) 12262--12270.

\bibitem{WHOblueprint2018}
https://www.who.int/blueprint/en/.

\bibitem{portillo2021epidemiological}
A.~Portillo, A.~M. Palomar, P.~Santib{\'a}{\~n}ez, J.~A. Oteo, Epidemiological
  aspects of {Crimean-Congo} hemorrhagic fever in {Western} {Europe}: what
  about the future?, Microorganisms 9~(3) (2021) 649.

\bibitem{fanelli2021risk}
A.~Fanelli, D.~Buonavoglia, Risk of {Crimean-Congo} haemorrhagic fever virus
  introduction and spread in {CCHF-free} countries in {Southern and Western
  Europe}: A semi-quantitative risk assessment, One Health 13 (2021) 100290.

\bibitem{mosquera1998evolution}
J.~Mosquera, F.~R. Adler, Evolution of virulence: a unified framework for
  co-infection and super-infection, Journal of Theoretical Biology 195~(3)
  (1998) 293--313.

\bibitem{allen2005asymptotic}
L.~J. Allen, N.~Kirupaharan, Asymptotic dynamics of deterministic and
  stochastic epidemic models with multiple pathogens, International Journal of
  Numerical Analysis and Modeling 2~(3) (2005) 329--344.

\bibitem{zhang2013co}
X.-S. Zhang, D.~De~Angelis, P.~J. White, A.~Charlett, R.~G. Pebody,
  J.~McCauley, Co-circulation of influenza a virus strains and emergence of
  pandemic via reassortment: the role of cross-immunity, Epidemics 5~(1) (2013)
  20--33.

\bibitem{slater2013modelling}
H.~C. Slater, M.~Gambhir, P.~E. Parham, E.~Michael, Modelling co-infection with
  malaria and lymphatic filariasis, PLoS Computational Biology 9~(6) (2013)
  e1003096.

\bibitem{yakob2013slaving}
L.~Yakob, G.~M. Williams, D.~J. Gray, K.~Halton, J.~A. Solon, A.~C. Clements,
  Slaving and release in co-infection control, Parasites \& Vectors 6 (2013)
  1--9.

\bibitem{lass2013generating}
S.~Lass, P.~J. Hudson, J.~Thakar, J.~Saric, E.~Harvill, R.~Albert, S.~E.
  Perkins, Generating super-shedders: co-infection increases bacterial load and
  egg production of a gastrointestinal helminth, Journal of the Royal Society
  Interface 10~(80) (2013) 20120588.

\bibitem{maliyoni2019stochastic}
M.~Maliyoni, F.~Chirove, H.~D. Gaff, K.~S. Govinder, A stochastic epidemic
  model for the dynamics of two pathogens in a single tick population,
  Theoretical Population Biology 127 (2019) 75--90, ($^{\ast}$).

\bibitem{cutler2021tick}
S.~J. Cutler, M.~Vayssier-Taussat, A.~Estrada-Pe{\~n}a, A.~Potkonjak, A.~D.
  Mihalca, H.~Zeller, Tick-borne diseases and co-infection: Current
  considerations, Ticks and tick-borne diseases 12~(1) (2021) 101607,
  ($^{\ast}$).

\bibitem{pabon2023bayesian}
F.~M. Pabon-Rodriguez, G.~D. Brown, B.~M. Scorza, C.~A. Petersen, Bayesian
  multivariate longitudinal model for immune responses to {Leishmania: A}
  tick-borne co-infection study, Statistics in Medicine (2023).

\bibitem{gao2016coinfection}
D.~Gao, T.~C. Porco, S.~Ruan, Coinfection dynamics of two diseases in a single
  host population, Journal of mathematical analysis and applications 442~(1)
  (2016) 171--188.

\bibitem{lou2017modelingco}
Y.~Lou, L.~Liu, D.~Gao, Modeling co-infection of {Ixodes} tick-borne pathogens,
  Mathematical Biosciences \& Engineering 14~(5\&6) (2017) 1301.

\bibitem{vogels2019arbovirus}
C.~B. Vogels, C.~R{\"u}ckert, S.~M. Cavany, T.~A. Perkins, G.~D. Ebel, N.~D.
  Grubaugh, Arbovirus coinfection and co-transmission: A neglected public
  health concern?, PLoS biology 17~(1) (2019) e3000130, ($^{\ast}$).

\bibitem{Cross2005}
P.~C. Cross, J.~O. Lloyd-Smith, P.~L.~F. Johnson, W.~M. Getz, Duelling
  timescales of host movement and disease recovery determine invasion of
  disease in structured populations, Ecology Letters 183 (2005) 587–595.
\newblock \href {https://doi.org/10.1111/j.1461-0248.2005.00760.x}
  {\path{doi:10.1111/j.1461-0248.2005.00760.x}}.

\bibitem{meehan2020probability}
M.~T. Meehan, R.~C. Cope, E.~S. McBryde, On the probability of strain invasion
  in endemic settings: accounting for individual heterogeneity and control in
  multi-strain dynamics, Journal of Theoretical Biology 487 (2020) 110109,
  ($^{\ast}$).

\bibitem{alizon2013multiple}
S.~Alizon, J.~C. De~Roode, Y.~Michalakis, Multiple infections and the evolution
  of virulence, Ecology Letters 16~(4) (2013) 556--567.

\bibitem{allen2019modelling}
L.~J. Allen, V.~A. Bokil, N.~J. Cunniffe, F.~M. Hamelin, F.~M. Hilker, M.~J.
  Jeger, Modelling vector transmission and epidemiology of co-infecting plant
  viruses, Viruses 11~(12) (2019) 1153, ($^{\ast}$).

\bibitem{white2019dynamics}
A.~White, E.~Schaefer, C.~W. Thompson, C.~M. Kribs, H.~Gaff, Dynamics of two
  pathogens in a single tick population, Letters in Biomathematics 6~(1) (2019)
  50, ($^{\ast}$).

\bibitem{bushman2019general}
M.~Bushman, R.~Antia, A general framework for modelling the impact of
  co-infections on pathogen evolution, Journal of the Royal Society Interface
  16~(155) (2019) 20190165.

\bibitem{kermack1927contribution}
W.~O. Kermack, A.~G. McKendrick, A contribution to the mathematical theory of
  epidemics, Proceedings of the Royal Society of London, Series A 115~(772)
  (1927) 700--721.

\bibitem{pfab2022time}
F.~Pfab, R.~M. Nisbet, C.~J. Briggs, A time-since-infection model for
  populations with two pathogens, Theoretical Population Biology 144 (2022)
  1--12.

\bibitem{rovenolt2022impact}
F.~H. Rovenolt, A.~T. Tate, The impact of coinfection dynamics on host
  competition and coexistence, The American Naturalist 199~(1) (2022) 91--107,
  ($^{\ast}$).

\bibitem{le2022disentangling}
T.~M.~T. Le, S.~Madec, E.~Gjini, Disentangling how multiple traits drive two
  strain frequencies in {SIS} dynamics with coinfection, Journal of Theoretical
  Biology 538 (2022) 111041, ($^{\ast}$).

\bibitem{saad2021superinfection}
C.~M. Saad-Roy, B.~T. Grenfell, S.~A. Levin, L.~Pellis, H.~B. Stage, P.~Van
  Den~Driessche, N.~S. Wingreen, Superinfection and the evolution of an initial
  asymptomatic stage, Royal Society Open Science 8~(1) (2021) 202212.

\bibitem{mclaughlin2022vector}
A.~A. McLaughlin, L.~Hanley-Bowdoin, G.~G. Kennedy, A.~L. Jacobson, Vector
  acquisition and co-inoculation of two plant viruses influences transmission,
  infection, and replication in new hosts, Scientific Reports 12~(1) (2022)
  20355.

\bibitem{chapwanya2021synergistic}
M.~Chapwanya, A.~Matusse, Y.~Dumont, On synergistic co-infection in crop
  diseases. the case of the {Maize Lethal Necrosis Disease}, Applied
  Mathematical Modelling 90 (2021) 912--942.

\bibitem{miller2022mathematical}
J.~Miller, T.~M. Burch-Smith, V.~V. Ganusov, Mathematical modeling suggests
  cooperation of plant-infecting viruses, Viruses 14~(4) (2022) 741.

\bibitem{lipsitch2009no}
M.~Lipsitch, C.~Colijn, T.~Cohen, W.~P. Hanage, C.~Fraser, No coexistence for
  free: neutral null models for multistrain pathogens, Epidemics 1~(1) (2009)
  2--13.

\bibitem{alizon2013co}
S.~Alizon, Co-infection and super-infection models in evolutionary
  epidemiology, Interface Focus 3~(6) (2013) 20130031, ($^{\ast\ast}$).

\bibitem{alizon2013parasite}
S.~Alizon, Parasite co-transmission and the evolutionary epidemiology of
  virulence, Evolution 67~(4) (2013) 921--933, ($^{\ast}$).

\bibitem{bhowmick2022ticks}
S.~Bhowmick, K.~K. Kasi, J.~Gethmann, S.~Fischer, F.~J. Conraths, I.~M.
  Sokolov, H.~H. Lentz, Ticks on the run: {A} mathematical model of
  {Crimean-Congo Haemorrhagic Fever (CCHF)} -- key factors for transmission,
  Epidemiologia 3~(1) (2022) 116--134, ($^{\ast}$).

\bibitem{gonzalez1992sexual}
J.-P. Gonzalez, J.-L. Camicas, J.-P. Cornet, O.~Faye, M.~Wilson, Sexual and
  transovarian transmission of {Crimean-Congo} haemorrhagic fever virus in
  {Hyalomma} truncatum ticks, Research in Virology 143 (1992) 23--28.

\bibitem{matser2009elasticity}
A.~Matser, N.~Hartemink, H.~Heesterbeek, A.~Galvani, S.~Davis, Elasticity
  analysis in epidemiology: an application to tick-borne infections, Ecology
  Letters 12~(12) (2009) 1298--1305.

\bibitem{gargili2017role}
A.~Gargili, A.~Estrada-Pe{\~n}a, J.~R. Spengler, A.~Lukashev, P.~A. Nuttall,
  D.~A. Bente, The role of ticks in the maintenance and transmission of
  {Crimean-Congo} hemorrhagic fever virus: {A} review of published field and
  laboratory studies, Antiviral esearch 144 (2017) 93--119.

\bibitem{spengler2019crimean}
J.~R. Spengler, {\'E}.~Bergeron, C.~F. Spiropoulou, {Crimean-Congo} hemorrhagic
  fever and expansion from endemic regions, Current Opinion in Virology 34
  (2019) 70--78.

\bibitem{gonzalez1998biological}
J.-P. Gonzalez, J.-L. Camicas, J.-P. Cornet, M.~Wilson, Biological and clinical
  responses of {West African sheep to Crimean-Congo} haemorrhagic fever virus
  experimental infection, Research in Virology 149~(6) (1998) 445--455.

\bibitem{hoch2018dynamic}
T.~Hoch, E.~Breton, Z.~Vatansever, Dynamic modeling of {Crimean-Congo}
  hemorrhagic fever virus (cchfv) spread to test control strategies, Journal of
  Medical Entomology 55~(5) (2018) 1124--1132, ($^{\ast}$).

\bibitem{van2002reproduction}
P.~Van~den Driessche, J.~Watmough, Reproduction numbers and sub-threshold
  endemic equilibria for compartmental models of disease transmission,
  Mathematical Biosciences 180~(1-2) (2002) 29--48.

\bibitem{diekmann1990definition}
O.~Diekmann, J.~A.~P. Heesterbeek, J.~A. Metz, On the definition and the
  computation of the basic reproduction ratio {$R_0$} in models for infectious
  diseases in heterogeneous populations, Journal of Mathematical Biology 28~(4)
  (1990) 365--382.

\bibitem{heesterbeek2007type}
J.~Heesterbeek, M.~Roberts, The type-reproduction number {T} in models for
  infectious disease control, Mathematical Biosciences 206~(1) (2007) 3--10.

\bibitem{johnstone2020incorporating}
S.~P. Johnstone-Robertson, M.~A. Diuk-Wasser, S.~A. Davis, Incorporating tick
  feeding behaviour into {$R_0$} for tick-borne pathogens, Theoretical
  Population Biology 131 (2020) 25--37, ($^{\ast\ast}$).

\bibitem{sorvillo2020towards}
T.~E. Sorvillo, S.~E. Rodriguez, P.~Hudson, M.~Carey, L.~L. Rodriguez, C.~F.
  Spiropoulou, B.~H. Bird, J.~R. Spengler, D.~A. Bente, Towards a sustainable
  one health approach to {Crimean-Congo} hemorrhagic fever prevention: Focus
  areas and gaps in knowledge, Tropical Medicine and Infectious Disease 5~(3)
  (2020) 113.

\bibitem{mpeshe2011mathematical}
S.~C. Mpeshe, H.~Haario, J.~M. Tchuenche, A mathematical model of {Rift Valley}
  fever with human host, Acta Biotheoretica 59~(3) (2011) 231--250.

\bibitem{belluccini2023stochastic}
G.~Belluccini, Stochastic models of cell population dynamics and tick-borne
  virus transmission, Ph.D. thesis, University of Leeds, ($^{\ast}$) (2023).

\bibitem{maliyoni2023multipatch}
M.~Maliyoni, H.~D. Gaff, K.~S. Govinder, F.~Chirove, Multipatch stochastic
  epidemic model for the dynamics of a tick-borne disease, Frontiers in Applied
  Mathematics and Statistics 9 (2023) 1122410, ($^{\ast}$).

\bibitem{artalejo2013exact}
J.~R. Artalejo, M.~J. Lopez-Herrero, On the exact measure of disease spread in
  stochastic epidemic models, Bulletin of Mathematical Biology 75 (2013)
  1031--1050.

\bibitem{yuan2011stochastic}
Y.~Yuan, L.~J. Allen, Stochastic models for virus and immune system dynamics,
  Mathematical Biosciences 234~(2) (2011) 84--94.

\bibitem{lin2023co}
D.~C.-D. Lin, S.-C. Weng, P.-N. Tsao, J.~J.~H. Chu, S.-H. Shiao, Co-infection
  of dengue and {Zika} viruses mutually enhances viral replication in the
  mosquito {Aedes aegypti}, Parasites \& Vectors 16~(1) (2023) 1--14,
  ($^{\ast}$).

\bibitem{cardano2007rules}
G.~Cardano, T.~R. Witmer, O.~Ore, The rules of algebra: {Ars Magna}, Vol. 685,
  Courier Corporation, 2007.

\bibitem{zwillinger2018crc}
D.~Zwillinger, {CRC} standard mathematical tables and formulas, CRC press,
  2018.

\end{thebibliography}

\end{document}